\documentclass[aps,preprint,showpacs,preprintnumbers,amsmath,amssymb]{revtex4}
\usepackage{graphicx}

\def\beq{\begin{equation}}
\def\eeq{\end{equation}}
\def\beqn{\begin{eqnarray}}
\def\eeqn{\end{eqnarray}}

\def\bseq{\begin{subequations}}
\def\eseq{\end{subequations}}

\def\1 {{\bf 1}}

\def\r {{\bf r}}

\def\d {{\bf d}}

\def\R {{\bf R}}
\def\C {{\bf C}}
\def\H {{\bf H}}
\def\U {{\bf U}}
\def\D {{\bf D}}
\def\T {{\bf T}}

\def\bcalH {\mbox{\boldmath $\mathcal H$}}
\def\brho {\mbox{\boldmath $\rho$}}

\begin{document}

\title{Many-body theory for systems with particle conversion:
Extending the multiconfigurational time-dependent Hartree method}

\author{Ofir E. Alon\footnote{E-mail: ofir@pci.uni-heidelberg.de},
        Alexej I. Streltsov\footnote{E-mail: alexej@pci.uni-heidelberg.de},
        and Lorenz S. Cederbaum\footnote{E-mail: lorenz.cederbaum@pci.uni-heidelberg.de}}

\affiliation{Theoretische Chemie, Physikalisch-Chemisches Institut, Universit\"at Heidelberg,\\
         Im Neuenheimer Feld 229, D-69120 Heidelberg, Germany}

\begin{abstract}
We derive a multiconfigurational 
time-dependent Hartree theory
for systems with particle conversion.
In such systems particles of one kind can convert to another kind 
and the total number of particles varies in time.
The theory thus extends the 
scope of the available and successful multiconfigurational
time-dependent Hartree methods -- 
which were solely 
formulated for and applied to systems with a fixed number of particles --
to new physical systems and problems.
As a guiding example we treat
explicitly a system where bosonic atoms can combine to 
form bosonic molecules and vise versa.
In the theory for particle conversion,
the time-dependent many-particle wavefunction 
is written as a sum of configurations made of a different number of particles,
and assembled from sets of atomic and molecular orbitals. 
Both the expansion coefficients and the orbitals forming the configurations
are time-dependent quantities that are fully determined according 
to the Dirac-Frenkel time-dependent variational principle. 
By employing the Lagrangian formulation of the Dirac-Frenkel variational principle 
we arrive at two sets of coupled equations of motion, 
one for the atomic and molecular orbitals and one for 
the expansion coefficients. 
The first set is comprised 
of first-order differential equations in time and 
nonlinear in-general integrodifferential equations in position space,
whereas the second set consists of first-order differential equations 
with coefficients forming a time-dependent Hermitian matrix. 
Particular attention is paid to the
reduced density matrices of 
the many-particle wavefunction that 
appear in the theory
and enter the equations of motion.
There are two kinds of reduced density matrices: 
particle-conserving reduced density matrices
which directly only couple configurations with the same 
number of atoms and molecules,
and particle non-conserving reduced density matrices 
which couple configurations with 
a different number of atoms and molecules.
Closed-form and compact equations of motion are derived 
for contact as well as general two-body interactions,
and their properties are 
analyzed and discussed.
\end{abstract}

\maketitle

\section{Introduction}\label{Intro}

The exploration of quantum dynamics of many-particle systems 
is a fundamental and on-going challenge of 
many branches in physics 
\cite{Book_dynamics1,Book_dynamics2,Nuclear_book,Book_dynamics3,Pit_Stri_book,Book_dynamics4}.
The equation of motion governing the evolution of quantum particles is,
in many cases, the well-known time-dependent Schr\"odinger equation.
Solving the time-dependent Schr\"odinger equation
for many-particle systems can rarely 
be made analytically or exactly,
which renders efficient 
approximations a must.

The multiconfigurational time-dependent Hartree method (MCTDH) \cite{CPL,JCP},
which has been developed in the past two decades,
is considered at present the most efficient
wave-packet propagation approach \cite{MCTDH_package} 
and has successfully and routinely been used for multi-dimensional 
dynamical systems consisting of distinguishable degrees-of-freedom,
such as molecular vibrations, 
see Ref.~\cite{JCP_24a,JCP_24b,PR,Dieter_review,Manthe_review,Lenz_CI,relaxation2,vib_new1,vib_new2,irene}.
The main idea behind the MCTDH method is to expand 
the time-dependent many-body wavefunction
of distinguishable particles by {\it time-dependent} configurations
that are assembled from {\it time-dependent} orbitals (one-body functions)
and optimized according to the Dirac-Frenkel variational principle \cite{DF1,DF2}.
In this way, 
a much larger effective subspace of the many-particle Hilbert space
can be spanned in practice in comparison to multiconfigurational 
expansions with {\it stationary} configurations.
By grouping several ``elementary'' degrees-of-freedom together and treating them
as ``generalized'' particles, 
the efficiency of the MCTDH algorithm increases \cite{JCP_24a,JCP_24b}.
Choosing to use MCTDH itself to propagate multi-dimensional
``generalized'' particles has led
to the idea of cascading \cite{Dieter_review}.
Finally,  
expanding the time-dependent orbitals 
themselves by other time-dependent orbitals,
and so on, 
putting the resulting time-dependent expansion 
under the Dirac-Frenkel variational principle,
leads to the multi-layer formulation of the MCTDH theory 
\cite{Multi_L} which further increases 
the efficiency of the MCTDH method for larger, 
complex systems.
The MCTDH can be applied to systems of identical particles.
In this direction,
we would like to mention that the MCTDH 
approach has very successfully been
employed to unveiling fundamental physics
of few-boson systems 
\cite{ZO_st1,ZO_st2,ZO_dy1,ZO_dy2,ZO_dy3,axel}
on the numerically-exact 
many-body level.

A new branch 
of MCTDH-based methods
has emerged after it had been realized that,
to effectively treat the dynamics of more than
a handful {\it identical} particles,
it is essential to use their quantum statistics,
Fermi-Dirac or Bose-Einstein, 
to eliminate the large amount of redundancies of coefficients
in the distinguishable-particle 
multiconfigurational expansion of the MCTDH wavefunction.
First, taking explicitly the antisymmetry of the many-fermion wavefunction 
to permutations of any two particles into account,
the fermionic version of MCTDH -- MCTDHF -- was independently 
developed by several groups \cite{MCTDHF1,MCTDHF2,MCTDHF3}. 
Shortly after, 
the bosonic version of MCTDH -- MCTDHB -- was developed in \cite{MCTDHB1,MCTDHB2}.
This advancement is in particular valuable since very-many bosons can reside in only 
a small number of orbitals owing to Bose-Einstein statistics,
thereby allowing the successful and quantitative 
attack of the dynamics of a much larger 
number of bosons with the MCTDHB theory.
For applications of MCTDHF to the many-body dynamics of 
at-present few-fermion systems with or without external laser field
see Refs.~\cite{applF1,applF2,applF3,applF4,applF5,applF6,applF7},
and for applications of MCTDHB for the many-body dynamics of 
repulsive and attractive bosonic systems 
Refs.~\cite{MCTDHB1,applB1,applB2,applB3}.
We mention that Ref.~\cite{applB3}
has combined optimal control theory
with MCTDHB. 

Five decades ago,
in his seminal paper,
L\"owdin defined the reduced density matrices
of many-fermion wavefunctions \cite{Lowdin}.
Since then,
reduced density matrices and, 
in particular, 
reduced two-body density matrices
is a fruitful and vivid research area 
including theory and applications in 
electronic structure of molecules, 
quantum phase-transitions,
and ground-state nuclear motion 
\cite{Slava,MAZZ1,MAZZ2,MAZZ3,MAZZ4,MAZZ5,MAZZ6}.
In the present context,
reduced one- and two-body density matrices
were first used to derive the stationary many-body states
within the general variational theory
with complete self-consistency for trapped bosonic systems --
the multiconfigurational Hartree for bosons (MCHB) \cite{MCHB}.
Later on,
the MCTDHF and MCTDHB were formulated in a unified manner,
making use of the reduced one- and two-body density matrices 
of the time-dependent many-body wavefunction \cite{Unified_paper}. 
Finally, treating mixtures of two kinds of identical particles in a unified manner,
and utilizing the reduced one- and two-body density matrices of the mixture's wavefunction,
a multiconfigurational time-dependent Hartree method 
for Fermi-Fermi (MCTDH-FF), 
Bose-Bose (MCTDH-BB) and 
Bose-Fermi (MCTDH-BF) mixtures
has been derived \cite{MIX}. 

The multiconfigurational time-dependent Hartree method and its versions
specified for identical particles and mixtures are 
{\it particle-conserving} many-body propagation theories.
Namely, 
they were solely 
formulated for  
and applied to systems with 
a fixed number of particles.
Conceptually,
they aim at describing systems of coupled degrees-of-freedom
or interacting particles which have first-quantization Hamiltonian.
This brings us to the theme of the present work,
which is to derive a multiconfigurational 
time-dependent Hartree theory
for systems with {\it particle conversion}.
In such systems particles of one kind can convert to another kind 
and the total number of particles varies in time.
Hence, 
they are generally represented by 
a phenomenological second-quantized
Hamiltonian which includes 
a {\it conversion term}.
Doing so,
we extend the scope of the available and successful multiconfigurational
time-dependent Hartree method and its versions
specified for identical particles and mixtures
to new physical systems and problems.
We abbreviate the multiconfigurational 
time-dependent Hartree theory
for systems with particle conversion
by MCTDH-{\it conversion}.

As a concrete and guiding example for a many-body system 
with particle conversion and without loss of generality,
we consider explicitly the conversion of 
bosonic atoms ($a$) 
to bosonic molecules ($m$)
via the `reaction' $2a \leftrightharpoons m$,
which has been a system of tremendous 
theoretical and experimental interest 
in quantum-gas physics 
\cite{PD1,2M1,Timmermans_review,Yurovsky,Timm1,EXP1,PD2,Beyond,Holland_PRL,Sadhan,Vardi,EXP2,EXP3,Jaksch,G_STOOF,Tilman,
EXP3h,EXP4,EXP5,Review_STOOF,QFT2,Holland_PRA,PRL_STOOF_Subir,EXP6,Meystre_Rev,2M2_2005,Band,Thorsten_Rev,Vardi_PRL,EXP7,NBIG}.

An effective quantum-field-theory-based Hamiltonian 
for atomic and molecular Bose-Einstein condensates (BECs)
coupled by conversion 
was first put forward by Drummond {\it et al.} in \cite{PD1}.
In \cite{2M1}, 
a proposition that a molecular BEC
could be produced by coherent photoassociation was made
and a phenomenological two-mode Hamiltonian to
describe this process was suggested.
A microscopic theory to derive the many-body Hamiltonian
of bosonic atoms and molecules with conversion 
and the respective Gross-Pitaevskii theory with conversion
were put forward in \cite{Timmermans_review},
also see \cite{Timm1}. 
In \cite{Yurovsky},
a coupled system of Gross-Pitaevskii equations
with conversion and deactivation-rate (dissipation) terms
has been derived. 
The validity of the two-mode approach for conversion,
at least in the homogeneous system,
was questioned in \cite{Beyond},
where dissociation of molecules to other 
than the
ground atomic mode 
signifies that one needs to
go beyond the two-mode approximation.
The importance of pair correlations 
in the 
dynamics of 
resonantly-coupled 
atomic and molecular BECs,
leading to significant deviations from the 
respective Gross-Pitaevskii theory,
was put forward in \cite{Holland_PRL}.
That even in the perfect two-mode limit
the mean-field theory with conversion can fail,
because of strong particle-particle entanglement
near the dynamically unstable molecular mode,
was reported in \cite{Vardi}.
A proposition to create a molecule BEC
from an atomic Mott-insulator phase with exactly 
two bosons per lattice 
site was made in \cite{Jaksch}.
A full microscopic theory to derive
the Hamiltonian of atoms and molecules with the conversion term
from the microscopic particle-conserving Hamiltonian
of a homogeneous gas of 
identical bosonic atoms with two internal states
was given in \cite{G_STOOF}, also see \cite{Review_STOOF}.
Quantum phase transitions and effects of rotations in homogeneous 
systems 
of atomic and molecular BECs with conversion have been
discovered in \cite{QFT2,PRL_STOOF_Subir} and \cite{Holland_PRA}, 
respectively.
Finally, in a (harmonic) trap,
confinement effects on the stimulated dissociation
(effective conversion rate)
of a molecular to an atomic BEC
were recently found in \cite{Vardi_PRL},
and unique phases (vortex configurations) 
of rotating interacting atomic-molecular
BECs in \cite{NBIG}.

Molecules were first produced from and identified
in a $^{\mathrm {87}}$Rb BEC 
by Wynar {\it et al.} \cite{EXP1}.
Soon after,
photoassociation of ultracold sodium molecules 
in an atomic 
BEC was made \cite{EXP2}.
Atomic-molecular coherence 
in a BEC [made of $^{\mathrm {85}}$Rb atoms]
was first achieved in \cite{EXP3}.
A pure molecular quantum gas produced from 
an atomic cesium BEC was reported in \cite{EXP3h},
and a quantum-degenerate gas of sodium molecules
in \cite{EXP4}.
More recently, 
with $^{\mathrm {87}}$Rb atoms 
in the Mott-phase of optical lattices,
state-selective conversion of atoms to molecules \cite{EXP6},
following the theoretical proposition in \cite{Jaksch},
and atom-molecule Rabi oscillations \cite{EXP7}
have been observed.

Finally,
systems with particle conversion 
can involve of course fermions, 
in the cold-atom world -- see the 
reviews \cite{Thorsten_Rev,F_Review} 
and references therein -- and beyond it.
In the latter context,
it is gratifying to 
mention 
the Friedberg-Lee model of superconductivity,
describing the conversion of two electrons to a single Cooper-pair
and vice versa by a boson-fermion Hamiltonian 
with a phenomenological conversion term \cite{SC0,SC1}. 

Let us return now to MCTDH-{\it conversion}, 
and put it in the particular context of 
interacting atomic and molecular BECs with conversion.
For the explicit scenario 
of the conversion `reaction' $2a{\leftrightharpoons}m$
dealt with throughout this work,
the theory shall be referred to as
MCTDH-[$2a{\leftrightharpoons}m$].
MCTDH-[$2a{\leftrightharpoons}m$],
as its particle-conserving predecessors 
\cite{CPL,JCP,PR,MCTDHF1,MCTDHF2,MCTDHF3,MCTDHB1,MCTDHB2,Unified_paper,MIX}, 
is intended for 
systems with 
a finite number of interacting particles,
typically trapped 
in an external potential.
As a first step, 
we extend or ``merge''
two theoretical approaches 
much in use in the literature:
the Gross-Pitaevskii theory with conversion
and the two-mode approximation,
see, e.g., 
Refs.~\cite{Timmermans_review,Timm1,PD2,Sadhan,Band,NBIG} and Refs.~\cite{2M1,Vardi,2M2_2005}, 
respectively. 
This results in
a fully variational theory where
the two modes -- the atomic and molecular orbitals --
and each and every expansion
coefficient in the two-mode many-body wavefunction
are fully optimized -- the orbitals in time and space
and the expansion coefficients in time -- 
according to the Dirac-Frenkel variational principle \cite{DF1,DF2}.
Our main aim 
is to go beyond any two-mode description
of the atomic-molecular coupled system
and present a fully-variational 
multiconfigurational
time-dependent many-body theory 
for bosonic atoms and molecules coupled by conversion
-- the MCTDH-[$2a{\leftrightharpoons}m$] theory.

The structure of the paper is as follows. 
We open in section \ref{secII} 
with the many-body Hamiltonian of the system
of atoms and molecules with conversion. 
In section \ref{two_mode_AM} 
we consider 
as mentioned above
a specific case of interest,
the fully-variational theory
where there are one atomic and one molecular orbitals.
This specific theory will be referred to as {\it conversion mean field}.
Next, 
section \ref{full_Hilbert}
is devoted to the general theory.
Both time-dependent as well as time-independent
theories are presented.
Finally, 
in section \ref{dis_sum} 
we put forward a summary 
and concluding remarks.
Complementary derivations and 
relevant matrix elements
are deferred to and collected 
in appendices 
\ref{Lagrange} and \ref{matrix_Appen},
respectively.

\section{The many-body Hamiltonian of interacting atoms and molecules with conversion}\label{secII}

As a concrete example for a many-body system with particle conversion and without loss of generality,
we consider a system of bosons which will be referred to as atoms ($a$)
and their conversion to another type of bosons which will be referred to as molecules ($m$)
via the `reaction' $2a \leftrightharpoons m$.
The many-body Hamiltonian of the coupled atom--molecule system is taken from the literature
of cold-atom physics 
\cite{PD1,Timmermans_review,G_STOOF,Holland_PRA} 
and is written for our needs as a sum of four terms:
\beq\label{ham_am_1}
\hat H^{(2a\rightleftharpoons m)} = \hat H^{(am)} + \hat W^{(2a\rightleftharpoons m)} =
\hat H^{(a)} + \hat H^{(m)} + \hat W^{(am)} + \hat W^{(2a\rightleftharpoons m)}.
\eeq
The first three terms are particle-conserving terms and together describe a mixture
of two kinds of interacting bosonic particles; $a$ (atoms) and $m$ (molecules):
\beqn\label{ham_am_2}
 \hat H^{(a)} &=& \hat h^{(a)} + \hat W^{(a)} = \nonumber \\
              &=& \int d\r \left[ \hat{\mathbf \Psi}^\dag_a(\r) \hat h^{(a)}(\r) \hat{\mathbf \Psi}_a(\r) +
 \frac{1}{2} \int d\r' \hat{\mathbf \Psi}^\dag_a(\r) \hat{\mathbf \Psi}^\dag_a(\r') \hat W^{(a)}(\r,\r') 
 \hat{\mathbf \Psi}_a(\r') \hat{\mathbf \Psi}_a(\r) \right], \nonumber \\
 \hat H^{(m)} &=& \hat h^{(m)} + \hat W^{(m)} = \nonumber \\
              &=& \int d\r \left[ \hat{\mathbf \Psi}^\dag_m(\r) \hat h^{(m)}(\r) \hat{\mathbf \Psi}_m(\r) +
  \frac{1}{2} \int d\r' \hat{\mathbf \Psi}^\dag_m(\r) \hat{\mathbf \Psi}^\dag_m(\r') \hat W^{(m)}(\r,\r') 
 \hat{\mathbf \Psi}_m(\r') \hat{\mathbf \Psi}_m(\r) \right], \nonumber \\
  \hat W^{(am)} &=&  \int d\r \int d\r' \hat{\mathbf \Psi}^\dag_a(\r) \hat{\mathbf \Psi}^\dag_m(\r') 
   \hat W^{(am)}(\r,\r') \hat{\mathbf \Psi}_m(\r') \hat{\mathbf \Psi}_a(\r). \
\eeqn
The last term 
describes the conversion of atoms to molecules and vise versa 
and is given by \cite{PD1,Timmermans_review,G_STOOF,Holland_PRA}:
\beqn\label{ham_am_3}
 \hat W^{(2a\rightleftharpoons m)} &=& \hat W^{(2a\rightharpoonup m)} + \hat W^{(m\rightharpoondown 2a)} = \nonumber \\
 &=& \frac{1}{\sqrt{2}} \int d\r \int d\r' 
\Bigg[\hat{\mathbf \Psi}^\dag_m\left(\frac{\r+\r'}{2}\right)
\hat W^{(2a\rightharpoonup m)}(\r,\r')\hat{\mathbf \Psi}_a(\r) \hat{\mathbf \Psi}_a(\r') + \nonumber \\
 &+& \hat{\mathbf \Psi}^\dag_a(\r') \hat{\mathbf \Psi}^\dag_a(\r) \hat W^{(m\rightharpoondown 2a)}(\r,\r')
 \hat{\mathbf \Psi}_m\left(\frac{\r+\r'}{2}\right)\Bigg], \nonumber \\
 & & \qquad \hat W^{(m\rightharpoondown 2a)}(\r,\r') = \left\{\hat W^{(2a\rightharpoonup m)}(\r,\r')\right\}^\dag. \
\eeqn
The coordinates entering the field operators in (\ref{ham_am_3}) represent the annihilation (creation) of two atoms,
one at position $\r$ the second at position $\r'$, and the creation (annihilation) of a molecule
at the center-of-mass coordinate $\R=\frac{\r+\r'}{2}$.
The atomic, molecular field operators satisfy the usual commutation relations for bosons:
$\left[\hat{\mathbf \Psi}_a(\r), \hat{\mathbf \Psi}^\dag_a(\r')\right] = 
\left[\hat{\mathbf \Psi}_m(\r), \hat{\mathbf \Psi}^\dag_m(\r')\right] = \delta(\r-\r')$
and  
$\left[\hat{\mathbf \Psi}_a(\r), \hat{\mathbf \Psi}_a(\r')\right] = 
\left[\hat{\mathbf \Psi}_m(\r), \hat{\mathbf \Psi}_m(\r')\right] = 0$.
Since the atoms and molecules are distinguishable, different particles,
their mutual field operators commute,
$\left[\hat{\mathbf \Psi}_a(\r), \hat{\mathbf \Psi}^\dag_m(\r')\right] = 
\left[\hat{\mathbf \Psi}_a(\r), \hat{\mathbf \Psi}_m(\r')\right] = 0$.
Finally, we note that the interaction terms appearing in the Hamiltonian
(\ref{ham_am_1}-\ref{ham_am_3}) are symmetric,
i.e.,
$\hat W^{(a)}(\r,\r')=\hat W^{(a)}(\r',\r),\ldots,\hat W^{(m\rightharpoondown 2a)}(\r,\r') = 
\hat W^{(m\rightharpoondown 2a)}(\r',\r)$,
because the Hamiltonian is symmetric to the exchange of position 
of any two particles of the same kind.

The Hamiltonian (\ref{ham_am_1}-\ref{ham_am_3}) commutes with the following particle-number operator
\beq\label{particle_number}
  \hat N = \hat N_a + 2\hat N_m = \int d\r \left[\hat{\mathbf \Psi}^\dag_a(\r)\hat{\mathbf \Psi}_a(\r) + 
 2\hat{\mathbf \Psi}^\dag_m(\r)\hat{\mathbf \Psi}_m(\r)\right],
\eeq
reflecting a conservation law in presence of particle conversion.
Accordingly, the Hilbert-space of the problem is a direct sum of Hilbert
subspaces with different number of atoms and molecules:
\{$N$ atoms; $0$ molecules\}$\oplus$\{$N-2$ atoms; $1$ molecule\}$\oplus$\{$N-4$ atoms; $2$ molecules\}$\oplus 
\ldots \oplus$\{$N-2\left[\frac{N}{2}\right]$ atoms; $\left[\frac{N}{2}\right]$ molecules\},
where $\left[j\right]$ means the greatest integer not exceeding $j$.

The purpose of this work is to treat the
many-body Hamiltonian with atom--molecule conversion (\ref{ham_am_1}-\ref{ham_am_3}) multiconfigurationally.
To this end, we expand the atomic $\hat{\mathbf \Psi}_a(\r)$ and 
molecular $\hat{\mathbf \Psi}_m(\r)$ field operators by two complete sets 
of {\it time-dependent} orbitals,
\beq\label{annihilation_def}
 \hat{\mathbf \Psi}_a(\r) = \sum_k \hat b_k(t)\phi_k(\r,t), \qquad 
 \hat{\mathbf \Psi}_m(\r) = \sum_{k'} \hat c_{k'}(t)\psi_{k'}(\r,t).
\eeq
The sets of atomic $\left\{\phi_k(\r,t)\right\}$
and molecular $\left\{\psi_{k'}(\r,t)\right\}$
orbitals span the 
{\it time-dependent}
Hilbert space in which the system is to be propagated.
The advantages of time-dependent multiconfigurational expansions, see the Introduction,
is the employment of {\it time-dependent} orbitals 
which change in time according to 
a time-dependent variational principle.
This allows one to use in practical computations
a smaller number of time-dependent orbitals
than the number of time-independent orbitals
that would have been required otherwise.
A general multiconfigurational expansion, 
see section \ref{full_Hilbert}, 
employs $M$ 
orbitals for the bosonic
atoms and $M'$ orbitals for the bosonic molecules.
In particular, 
even if only one orbital is available for the bosonic atoms
and another one for the bosonic molecules, 
the resulting theory
{\it goes beyond} 
the Gross-Pitaevskii 
mean-field 
theory for this system \cite{Timmermans_review},
see subsequent section \ref{two_mode_AM}.

Finally, 
it is convenient to derive the relevant results first
for the popular contact interaction,
\beqn\label{contact_potentials}
 & & \hat W^{(a)}(\r,\r') = \lambda_a \delta(\r-\r'), \qquad \hat W^{(m)}(\r,\r') = \lambda_m \delta(\r-\r'), \nonumber \\
 & & \!\!\!\!\!\!\!\! \hat W^{(am)}(\r,\r') = \lambda_{am} \delta(\r-\r'), 
\qquad \hat W^{(2a\rightharpoonup m)}(\r,\r') = \hat W^{(m\rightharpoondown 2a)}(\r,\r') 
= \lambda_{con} \delta(\r-\r'). \
\eeqn
Thus, 
substituting Eqs.~(\ref{annihilation_def},\ref{contact_potentials})
into the generic Hamiltonian (\ref{ham_am_1}-\ref{ham_am_3}) we get
\beqn\label{ham_contact_inter}
& & 
 \hat H^{(2a\rightleftharpoons m)} = 
\sum_{k,q} \left<\phi_k\left|\hat h^{(a)}\right|\phi_q\right> \hat b_k^\dag \hat b_q + 
\frac{\lambda_a}{2}\sum_{k,s,l,q} \left<\phi_k\phi_s|\phi_q\phi_l\right> 
\hat b_k^\dag \hat b_s^\dag \hat b_l \hat b_q + \nonumber \\
 & & + \sum_{k',q'} \left<\psi_{k'}\left|\hat h^{(m)}\right|\psi_{q'}\right> \hat c_{k'}^\dag \hat c_{q'} + 
 \frac{\lambda_m}{2} \sum_{k',s',l',q'} \left<\phi_{k'}\phi_{s'}|\phi_{q'}\phi_{l'}\right>  
\hat c_{k'}^\dag \hat c_{s'}^\dag \hat c_{l'} \hat c_{q'} + \\
& & + \lambda_{am}\sum_{k,k',q,q'} \left<\phi_k\psi_{k'}|\phi_q\psi_{q'}\right>
 \hat b_k^\dag \hat b_q \hat c_{k'}^\dag \hat c_{q'} + 
 \frac{\lambda_{con}}{\sqrt{2}} 
\sum_{k',k,q} \left[ 
\left<\psi_{k'}\left|\right.\phi_k \phi_q\right> \hat c_{k'}^\dag \hat b_k \hat b_q +
\left<\phi_q \phi_k\left|\right.\psi_{k'}\right> \hat b_q^\dag \hat b_k^\dag \hat c_{k'}
\right]. \nonumber \
\eeqn
Here and hereafter,
the dependence of quantities on time is not shown 
explicitly whenever unambiguous.
Below, we will work throughout 
sections \ref{two_mode_AM} and \ref{contact}
with the contact-interaction Hamiltonian (\ref{ham_contact_inter})
and handle the case of general interactions 
(\ref{ham_am_1}-\ref{ham_am_3}) thereafter,
in section \ref{non_contact}. 

\section{The simplest case of atom--molecule conversion: Conversion mean field
 (Fully-variational two-mode approximation)}\label{two_mode_AM} 

\subsection{The multiconfigurational ansatz}\label{two_mode_AM_asnatz}

To introduce the nomenclature in the first stage of this work 
and, independently, 
as an interesting and relevant problem for itself,
we consider the resulting theory when there is only
one orbital available
for the (bosonic) atoms and one orbital available for the (bosonic) molecules.
The atomic orbital will be denoted by $\phi_1(\r,t)\equiv\phi_a(\r,t)$ and the molecular orbital
by $\psi_1(\r,t)\equiv\psi_m(\r,t)$.
The corresponding creation operators are denoted by 
$\hat b_1^\dag(t)\equiv\hat b_a^\dag(t)$ and $\hat c_1^\dag(t)\equiv\hat c_m^\dag(t)$.
The atomic and molecular creation, annihilation operators obey
the bosonic commutation relations corresponding to the field operators. 

The problem we wish to solve may now be formulated.
In the present section we would like to derive a multiconfigurational 
theory for atom--molecule conversion 
which is exact in the smallest Hilbert subspace 
possible for bosonic species,
namely, 
the Hilbert space spanned by 
the {\it single} molecular orbital 
$\psi_m(\r,t)$ and {\it single} atomic 
orbital $\phi_a(\r,t)$. 
We term this 
specific case of 
the general theory: {\it conversion mean field}.
More technically,
this theory
is a fully-variational extension of 
the literature two-mode approximation \cite{2M1,Vardi}
and,
of course, 
of the Gross-Pitaevskii 
theory 
with conversion \cite{Timmermans_review}.

The multiconfigurational wavefunction takes on the following form:
\beqn\label{2orb_Phi}
& & \left|\Psi(t)\right> = \sum_{p=0}^{\left[N/2\right]} C_p(t) \left|N-2p,p;t\right>, \nonumber \\
& & \left|N-2p,p;t\right> = \frac{1}{\sqrt{(N-2p)! p!}} 
\left(\hat b_a^\dag(t)\right)^{N-2p} \left(\hat c_m^\dag(t)\right)^p \left|vac\right>, \
\eeqn
where $\left|vac\right>$ is a {\it common} vacuum of no atoms and no molecules.
The index $p$ enumerates the number of bosonic
molecules in the system.
The corresponding number of atoms is $N-2p$.
$N$ is the maximal number of atoms in the system
which is obtained when there are no molecules.
Obviously,
$\left|\Psi(t)\right>$
is an eigenfunction of the particle-number operator $\hat N$,
Eq.~(\ref{particle_number}),
with the eigenvalue $N$.
The atomic and molecular number operators in the relevant Hilbert space boil
down to $\hat N_a=\hat b_a^\dag \hat b_a$ and $\hat N_m=\hat c_m^\dag \hat c_m$,
respectively.
The size of this
Hilbert space is $\left[\frac{N}{2}\right]+1$.

\subsection{The functional action $S$ of the time-dependent Schr\"odinger equation
and its evaluation}\label{two_mode_AM_energy}

Solving the time-dependent Schr\"odinger equation with the Hamiltonian
(\ref{ham_contact_inter}) and the multiconfigurational
ansatz (\ref{2orb_Phi}) means finding
the equations governing the time evolution of 
the atomic and molecular orbitals,
$\phi_a(\r,t)$ and $\psi_m(\r,t)$, 
and of the expansion coefficients $\{C_p(t)\}$.
The derivation of these equations of motion for 
$\phi_a(\r,t)$, $\psi_m(\r,t)$, and $\{C_p(t)\}$
requires a time-dependent variational principle.
We employ 
the Lagrangian formulation of the (Dirac-Frenkel) 
time-dependent variational principle \cite{LF1,LF2},
also see Refs.~\cite{MCTDHB2,Unified_paper,MIX},
and write the functional action of the time-dependent 
Schr\"odinger equation which takes on the form:
\beqn\label{action_functional_AM}
 & & S\left[\{C_p(t)\},\phi_a(\r,t),\psi_m(\r,t)\right] =  
 \int dt \Bigg\{\left<\Psi(t)\left| \hat H^{(2a\leftrightharpoons m)} -
i\frac{\partial}{\partial t} \right|\Psi(t)\right> \nonumber \\
 & & - \mu_a(t)\left[\left<\phi_a(\r,t)|\phi_a(\r,t)\right> - 1 \right] 
- \mu_m(t)\left[\left<\psi_m(\r,t)|\psi_m(\r,t)\right> - 1 \right] \nonumber \\
 & & - \varepsilon(t)\left[\sum_{p=0}^{\left[N/2\right]} \left|C_p(t)\right|^2 - 1\right]\Bigg\}. \ 
\eeqn
The time-dependent Lagrange multiplies $\mu_a(t)$, $\mu_m(t)$ and $\varepsilon(t)$
are introduced to ensure normalization of the atomic $\phi_a(\r,t)$ and molecular $\psi_m(\r,t)$ orbitals 
and of the expansion coefficients $\{C_p(t)\}$ at all times.
$\mu_a(t)$ and $\mu_m(t)$ also serve another role.
They exactly ``compensate''
for those terms
appearing within the Dirac-Frenkel 
formulation of the variational principle 
$\left<\delta \Psi(t)\left| \hat H^{(2a\leftrightharpoons m)} -i\frac{\partial}{\partial t} \right|\Psi(t)\right>$ 
\cite{DF1,DF2},
i.e., when the variation of $\Psi(t)$ is performed before 
the expectation value 
$\left<\Psi(t)\left| \hat H^{(2a\leftrightharpoons m)} -i\frac{\partial}{\partial t} \right|\Psi(t)\right>$
is evaluated;
see in this context \cite{MCTDHB2,LF2}.
We shall 
see below and 
more elaborately 
in appendix \ref{Lagrange_two_mode} 
that these Lagrange multipliers can be eliminated from the 
resulting equations of motion by making use of the normalization 
of the orbitals in combination with 
unitary transformations.

The expectation value of the Hamiltonian in Eq.~(\ref{action_functional_AM}) 
can be expressed in two equivalent forms. 
The first form depends explicitly on the orbitals
$\phi_a(\r,t)$, $\psi_m(\r,t)$ and the second on the expansion coefficients $\{C_p(t)\}$.
The two forms are needed to derive the respective equations of motion for the
orbitals and expansion coefficients.

\subsubsection{Orbital-explicit expression of $S$}\label{two_mode_AM_energy_orbital}

Utilizing the multiconfigurational expansion 
(\ref{2orb_Phi})
and the individual terms of the 
many-body Hamiltonian
(\ref{ham_contact_inter}),
the first form 
of the expectation value of $\hat H^{(a\leftrightharpoons m)} - i\frac{\partial}{\partial t}$ reads:
\beqn\label{Hamiltonian_matrix_element_AM_a}
& & \!\!\!\!\!\!\!\!\! \left<\Psi(t)\left| \hat H^{(2a\leftrightharpoons m)} 
- i\frac{\partial}{\partial t}\right|\Psi(t)\right> =
\left<\hat N_a\right> \left<\phi_a\left|\hat h^{(a)} - i\frac{\partial}{\partial t} \right|\phi_a\right> + \nonumber \\
& & \!\!\!\!\!\!\! + \frac{\lambda_a}{2} \left<\hat N_a(\hat N_a-1)\right> \left<\phi_a^2\left|\right.\phi_a^2\right> + 
\left<\hat N_m\right> \left<\psi_m\left|\hat h^{(m)} - i\frac{\partial}{\partial t}\right|\psi_m\right> + \nonumber \\
& & \!\!\!\!\!\!\! + \frac{\lambda_m}{2} \left<\hat N_m(\hat N_m-1)\right> \left<\psi_m^2\left|\right.\psi_m^2\right> + 
 \lambda_{am} \left<\hat N_a \hat N_m\right> \left<\phi_a\psi_m\left|\right.\phi_a\psi_m\right> + \nonumber \\
& & \!\!\!\!\!\!\! + \frac{\lambda_{con}}{\sqrt{2}} \left[\left<\hat c_m^\dag \hat b_a \hat b_a\right> 
\left<\psi_m\left|\right.\phi_a^2\right> +
\left<\hat b_a^\dag \hat b_a^\dag \hat c_m\right> \left<\phi_a^2\left|\right.\psi_m\right>
\right] - i  \sum_{p=0}^{\left[N/2\right]} C_p^\ast \frac{\partial C_p}{\partial t}. \
\eeqn
In Eq.~(\ref{Hamiltonian_matrix_element_AM_a})
and hereafter we use the shorthand notation for expectation values of operators with respect to $\Psi(t)$: 
$\left<\hat N_a\right> \equiv \left<\Psi(t)\left|\hat N_a\right|\Psi(t)\right>$,  
$\left<\hat b_a^\dag \hat b_a^\dag \hat c_m\right> \equiv 
\left<\Psi(t)\left|\hat b_a^\dag \hat b_a^\dag \hat c_m\right|\Psi(t)\right>$, etc.
We can indeed see that 
expression (\ref{Hamiltonian_matrix_element_AM_a}) 
depends explicitly on the orbitals $\phi_a(\r,t)$ and $\psi_m(\r,t)$
through integrals over one-body terms, 
two-body interaction terms,
and the conversion term: 
$\left<\phi_a\left|\hat h^{(a)} - i\frac{\partial}{\partial t} \right|\phi_a\right>$, 
$\left<\phi_a\psi_m\left|\right.\phi_a\psi_m\right>$, etc. 

Making use of the multiconfigurational expansion (\ref{2orb_Phi}),
we can express the above expectation values  
$\left<\Psi(t)\left| \ldots \right|\Psi(t)\right>$ 
in a closed form.
The expectation values of particle-conserving operators, 
like the number operators $\hat N_a$ and $\hat N_m$, with respect to $\Psi(t)$ read:
\beqn\label{matrix_elements_fock1_conserve}
& & \!\!\!\! \left<\hat N_a\right> = \sum_{p=0}^{\left[N/2\right]} (N-2p) \left|C_p(t)\right|^2, \qquad 
  \left<\hat N_m\right> = \sum_{p=0}^{\left[N/2\right]} p \left|C_p(t)\right|^2, \nonumber \\ 
& & \!\!\!\! \left<\hat N_a(\hat N_a-1)\right> = \sum_{p=0}^{\left[N/2\right]} (N-2p)(N-2p-1) \left|C_p(t)\right|^2, \nonumber \\
& & \!\!\!\! \left<\hat N_m(\hat N_m-1)\right> = \sum_{p=0}^{\left[N/2\right]} p(p-1) \left|C_p(t)\right|^2, \qquad 
 \left<\hat N_a \hat N_m\right> = \sum_{p=0}^{\left[N/2\right]} p(N-2p) \left|C_p(t)\right|^2. \
\eeqn
We see that the
dependence 
of the expectation values (\ref{matrix_elements_fock1_conserve})
on the expansion coefficients is only
through weighted sums $\sum_{p=0}^{\left[N/2\right]}$ 
of the terms $\left|C_p(t)\right|^2$,
i.e., that configurations of a different
number $p$ of molecules are not directly coupled.
The expectation values of particle non-conserving operators,
originating from the conversion of particles, are given by
\beqn\label{matrix_elements_fock1_non_conserve}
& & \left<\hat c_m^\dag \hat b_a \hat b_a\right> = \sum_{p=1}^{\left[N/2\right]} \sqrt{p(N-2p+1)(N-2p+2)}
 \, C_p^\ast(t)C_{p-1}(t), \nonumber \\
& & \left<\hat b_a^\dag \hat b_a^\dag \hat c_m\right> = 
\left\{\left<\hat c_m^\dag \hat b_a \hat b_a\right>\right\}^\ast, \
\eeqn
and seen to couple directly 
configurations with a different number of 
$p$ and $p-1$ molecules.

\subsubsection{Expansion-coefficient-explicit expression of $S$}\label{two_mode_AM_energy_coeff}

Utilizing the multiconfigurational expansion 
(\ref{2orb_Phi})
and the many-body Hamiltonian (\ref{ham_contact_inter}) as a whole,
we can express 
the expectation value of $\hat H^{(2a\leftrightharpoons m)} - i\frac{\partial}{\partial t}$
in the functional action (\ref{action_functional_AM}) as an explicit function of the 
expansion coefficients $\{C_p(t)\}$. 
One readily finds,
\beqn\label{Hamiltonian_matrix_element_AM_b}
& & \!\!\!\!\!\!\!\!\!
\left<\Psi(t)\left| \hat H^{(2a\leftrightharpoons m)} - i\frac{\partial}{\partial t}\right|\Psi(t)\right> = \nonumber \\
& & \!\!\!\!\!\!\!\!\!
 = \sum_{p=0}^{\left[N/2\right]} C_p^\ast \left[\sum_{p'=0}^{\left[N/2\right]}  
\left<N-2p,p;t\left|\hat H^{(2a\leftrightharpoons m)} - i\frac{\partial}{\partial t}\right|N-2p',p';t\right> C_{p'} 
- i \frac{\partial C_p}{\partial t}\right]. \
\eeqn
Eq.~(\ref{Hamiltonian_matrix_element_AM_b}) 
contains yet another type of matrix elements,
which are the representation of $\hat H^{(2a\leftrightharpoons m)} - i\frac{\partial}{\partial t}$
in the subspace of configurations $\left\{\left|N-2p,p;t\right>\right\}$.
These matrix elements can be 
evaluated explicitly.
We divide them into two types,
recalling that the Hamiltonian is expressed 
as a sum of particle-conserving and particle non-conserving parts,
$\hat H^{(2a\rightleftharpoons m)} = \hat H^{(am)} + \hat W^{(2a\rightleftharpoons m)}$.
The diagonal, 
or particle-conserving matrix elements read:
\beqn\label{matrix_elements_Hamil_diag}
 & & \!\!\!\!\!\!\!\!\!\! 
\left<N-2p,p;t\left|\hat H^{(am)} - i\frac{\partial}{\partial t}\right|N-2p,p;t\right> = 
 (N-2p)\left<\phi_a\left|\hat h^{(a)} - i\frac{\partial}{\partial t} \right|\phi_a\right> + \nonumber \\
& & + \frac{\lambda_a}{2} (N-2p)(N-2p-1) \left<\phi_a^2\left|\right.\phi_a^2\right> 
+ p \left<\phi_m\left|\hat h^{(m)} - i\frac{\partial}{\partial t}\right|\phi_m\right> + \nonumber \\
& & + \frac{\lambda_m}{2} p(p-1) \left<\phi_m^2\left|\right.\phi_m^2\right> +
\lambda_{am} p(N-2p) \left<\phi_a\phi_m\left|\right.\phi_a\phi_m\right>, \qquad p=0,\ldots,[N/2]. \
\eeqn
The off-diagonal, 
particle non-conserving matrix elements,
originating from the conversion term 
$\hat W^{(2a\rightleftharpoons m)}=\hat W^{(2a\rightharpoonup m)} + \hat W^{(m\rightharpoondown 2a)}$,
take on the following form:
\beqn\label{matrix_elements_Hamil_off_diag}
& & \!\!\!\!\!\!\!\!\!\! 
\left<N-2p,p;t\left|\hat W^{(2a\rightharpoonup m)}\right|N-2(p-1),p-1;t\right> = \nonumber \\
& & = \frac{\lambda_{con}}{\sqrt{2}}
\sqrt{p(N-2p+1)(N-2p+2)} \left<\phi_m\left|\right.\phi_a^2\right>, \qquad  p=1,\ldots,[N/2], \nonumber \\
& & \!\!\!\!\!\!\!\!\!\!
 \left<N-2(p-1),p-1;t\left|\hat W^{(m\rightharpoondown 2a)}\right|N-2p,p;t\right> = \nonumber \\
& & = \left\{\left<N-2p,p;t\left|\hat W^{(2a\rightharpoonup m)}\right|N-2(p-1),p-1;t\right>\right\}^\ast,
 \qquad  p=1,\ldots,[N/2]. \
\eeqn
All other matrix elements of $\hat H^{(2a\leftrightharpoons m)} - i\frac{\partial}{\partial t}$
in the subspace of configurations $\left\{\left|N-2p,p;t\right>\right\}$ vanish.
With explicit expressions of the functional action (\ref{action_functional_AM})
we can now proceed and derive the equations of motion of $\Psi(t)$.

\subsection{The equations of motion for $\Psi(t)$}\label{two_mode_AM_EOM}

We perform the variation of 
the action functional 
(\ref{action_functional_AM},\ref{Hamiltonian_matrix_element_AM_a})
with respect to the orbitals and coefficients. 
Equating the variation of $S\left[\{C_p(t)\},\phi_a(\r,t),\psi_m(\r,t)\right]$
with respect to the orbitals to zero,
eliminating the Lagrange multipliers $\mu_a(t)$ and $\mu_m(t)$ from the resulting equations
(see appendix \ref{Lagrange_two_mode} for details),
and dividing the result by $\left<\hat N_a\right>$ and $\left<\hat N_m\right>$ respectively,
we obtained the following equations of motion for the orbitals:
\beqn\label{simplest_EOM_orbitals}
& & \hat {\mathbf P}^{(a)} i\left|\dot \phi_a\right> = 
\hat {\mathbf P}^{(a)} 
\left\{\left[\hat h^{(a)} + \Lambda_a(t) |\phi_a|^2 + \Lambda_{am}(t) |\psi_m|^2 \right] \left|\phi_a\right> + 
 \sqrt{2} \Lambda_{con}(t) \phi_a^\ast \left|\psi_m\right>\right\}, \nonumber \\
& & \hat {\mathbf P}^{(m)} i\left|\dot \psi_m\right>  =
\hat {\mathbf P}^{(m)} 
 \left\{\left[\hat h^{(m)} + \Lambda_m(t) |\psi_m|^2 + \Lambda_{ma}(t) |\phi_a|^2 \right] \left|\psi_m\right> + 
\frac{\Lambda'_{con}(t)}{\sqrt{2}} \phi_a \left|\phi_a\right>\right\}, \
\eeqn
where the shorthand notation 
$\dot \phi_a \equiv \frac{\partial \phi_a}{\partial t}$,
$\dot \psi_m \equiv \frac{\partial \psi_m}{\partial t}$
is used here and hereafter.
The ``interaction strengths'' are given by
\beqn\label{time_dependent_interaction_str}
 & & \!\!\!\!\!\!\!\!\!\!\!\! 
\Lambda_a(t)=\lambda_a\frac{\left<\hat N_a(\hat N_a-1)\right>}{\left<\hat N_a\right>}, \ \  
\Lambda_{am}(t)=\lambda_{am} \frac{\left<\hat N_a \hat N_m\right>}{\left<\hat N_a\right>}, \ \  
\Lambda_{con}(t) = \lambda_{con} \frac{\left<\hat b_a^\dag \hat b_a^\dag \hat c_m\right>}
{\left<\hat N_a\right>}, \nonumber \\
 & &  \!\!\!\!\!\!\!\!\!\!\!\!
\Lambda_m(t)=\lambda_m \frac{\left<\hat N_m(\hat N_m-1)\right>}{\left<\hat N_m\right>}, \ \
\Lambda_{ma}(t) = \lambda_{am} \frac{\left<\hat N_a \hat N_m\right>}{\left<\hat N_m\right>}, \ \
\Lambda'_{con}(t) = \lambda_{con}\frac{\left<\hat c_m^\dag \hat b_a \hat b_a\right>}{\left<\hat N_m\right>}
\eeqn 
and {\it vary in time} 
due to the conversion
of atoms to molecules 
and vise versa.

The quantities appearing on both the right- and left-hand sides of 
equations of motion (\ref{simplest_EOM_orbitals})
are projection operators and given by
\beq\label{projection_conver_coherent}
\hat {\mathbf P}^{(a)} = 1 - \left|\phi_a\left>\right<\phi_{a}\right|, \qquad
\hat {\mathbf P}^{(m)} = 1 - \left|\psi_m\left>\right<\psi_{m}\right|.
\eeq
When acting on one-body functions in the atomic and molecular spaces,
$\hat {\mathbf P}^{(a)}$ and $\hat {\mathbf P}^{(m)}$
project these functions 
onto the subspaces
orthogonal to 
the orbitals
$\phi_a(\r,t)$ and $\psi_m(\r,t)$,
respectively.
The projection operators 
$\hat {\mathbf P}^{(a)}$, $\hat {\mathbf P}^{(m)}$
emerge when one eliminates the Lagrange multipliers
$\mu_a(t)$, $\mu_m(t)$
from the equations of motion, 
see appendix \ref{Lagrange_two_mode}.

The appearance of the projection operator on both
the left- and right-hand sides
of Eq.~(\ref{simplest_EOM_orbitals})
makes (\ref{simplest_EOM_orbitals}) 
a cumbersome set of two coupled integrodifferential non-linear equations.
Can one simplify the matters? The answer is positive.

To this end,
we invoke 
the invariance properties
of the many-particle wavefunction $\Psi(t)$.
Specifically, 
we can multiply the atomic $\phi_a(\r,t)$ and molecular $\psi_m(\r,t)$
orbitals
by time-dependent phase factors to give transformed orbitals 
$\overline{\phi}_a(\r,t)$ and $\overline{\psi}_m(\r,t)$,
and from the latter assemble 
transformed configurations
$\left|N-2p,p;t\right> \to \overline{\left|N-2p,p;t\right>}$.
Then,
we can compensate for transforming the orbitals
by the ``reverse'' transformation 
of the expansion coefficients
$\{C_p(t)\} \to \{\overline{C}_p(t)\}$.
Overall,
we write this 
invariance of the many-body wavefunction as follows:
\beq\label{trans_appen_Psi_text}
 \left|\Psi(t)\right> = \sum_{p=0}^{\left[N/2\right]} C_p(t) \left|N-2p,p;t\right> =
 \sum_{p=0}^{\left[N/2\right]} \overline{C}_p(t) \overline{\left|N-2p,p;t\right>}. \
\eeq 
Clearly,
unitary transformations of the orbitals 
and the respective 
transformation of the expansion coefficients
neither change the size of the Hilbert space
nor couple configurations 
with a different number of molecules.
To express these properties
we use the same summation index $p$
in both the middle part and right-hand side
of Eq.~(\ref{trans_appen_Psi_text}).

We can now make use of the invariance relation (\ref{trans_appen_Psi_text})
to simplify Eq.~(\ref{simplest_EOM_orbitals}).
Specifically,
there exists one unitary transformation that 
eliminates the projection operators
acting on the time-derivatives (left-hand sides)
in Eq.~(\ref{simplest_EOM_orbitals}),
without introducing any further constraint 
into 
the equations of motion; 
see appendix \ref{Lagrange_two_mode} for more details.
The equations of motion for the
atomic and molecular orbitals thus 
finally read:
\beqn\label{simplest_EOM_orbitals_final}
& & i\left|\dot \phi_a\right> = 
\hat {\mathbf P}^{(a)} 
\left\{\left[\hat h^{(a)} + \Lambda_a(t) |\phi_a|^2 + \Lambda_{am}(t) |\psi_m|^2 \right] \left|\phi_a\right> + 
 \sqrt{2} \Lambda_{con}(t) \phi_a^\ast \left|\psi_m\right>\right\}, \nonumber \\
& & i\left|\dot \psi_m\right>  =
\hat {\mathbf P}^{(m)} 
 \left\{\left[\hat h^{(m)} + \Lambda_m(t) |\psi_m|^2 + \Lambda_{ma}(t) |\phi_a|^2 \right] \left|\psi_m\right> + 
\frac{\Lambda'_{con}(t)}{\sqrt{2}} \phi_a \left|\phi_a\right>\right\}. \
\eeqn
Eq.~(\ref{simplest_EOM_orbitals_final}) 
has the following property.
Operating from the left with $\left<\phi_a\right|$ and $\left<\psi_m\right|$,
respectively,
we obtain the relations
\beq\label{MCTDHI_mix_conv_const_coherent}
\left<\phi_a\left|\right.\dot\phi_a\right> = 0, \qquad
\left<\psi_m\left|\right.\dot\psi_m\right> = 0,
\eeq
clearly ensuring that initially-normalized
orbitals remain normalized for 
all times.
We can see the meaning of the unitary transformation
carrying Eq.~(\ref{simplest_EOM_orbitals})
to Eq.~(\ref{simplest_EOM_orbitals_final}).
This unitary 
transformation takes normalized time-dependent orbitals,
$\left<\phi_a\left|\right.\phi_a\right>=1$ and $\left<\psi_m\left|\right.\psi_m\right>=1$,
which therefore satisfy the general relations
$\frac{\partial \left<\phi_a\left|\right.\phi_a\right>}{\partial t} = 
\left<\dot\phi_a\left|\right.\phi_a\right> + \left<\phi_a\left|\right.\dot\phi_a\right> = 0$
and
$\frac{\partial \left<\psi_m\left|\right.\psi_m\right>}{\partial t} = 
\left<\dot\psi_m\left|\right.\psi_m\right> + \left<\psi_m\left|\right.\dot\psi_m\right> = 0$,
and transforms them to
time-dependent orbitals
satisfying the specific differential condition 
(\ref{MCTDHI_mix_conv_const_coherent}).

Before we move to the corresponding working equations for the expansion
coefficients $\{C_p(t)\}$, 
it is instructive to enquire
whether we could further simplify the 
equations of motion (\ref{simplest_EOM_orbitals_final}),
by eliminating the projection operators 
$\hat {\mathbf P}^{(a)}$, $\hat {\mathbf P}^{(m)}$ 
also from the right-hand sides.
The answer is in general negative.
If we could eliminate the projection operators remaining
on the right-hand sides,
it means that conditions (\ref{MCTDHI_mix_conv_const_coherent}) 
are not satisfied any more.
What would then guarantee that
the atomic and molecular orbitals remain
normalized at all times?
It turns out 
that the condition for that is:
$\mathrm{Im}\left\{\lambda_{con} 
\left<c_m^\dag b_a b_a\right> \left<\psi_m\left|\right.\phi_a\phi_a\right>\right\} = 0$
for all times.
In turn,
even if this condition is satisfied at $t=0$,
it is not in general guaranteed that it remains so for all times.
Thus, the {\it presence of particle conversion} 
does not allow one to eliminate the
projection operators $\hat {\mathbf P}^{(a)}$, $\hat {\mathbf P}^{(m)}$ 
also from the right-hand sides of the equations of motion 
(\ref{simplest_EOM_orbitals_final}).
Alternatively speaking,
in the absence of particle conversion,
it is possible to eliminate the projection operators 
completely from (\ref{simplest_EOM_orbitals_final}),
see in this context \cite{TDMF}.

To derive the equations of motion of $\{C_p(t)\}$,
we equate the variation of the 
action functional (\ref{action_functional_AM},\ref{Hamiltonian_matrix_element_AM_b})
with respect to the expansion coefficients to zero
and eliminate 
the Lagrange multiplier $\varepsilon(t)$ 
(see for details appendix \ref{Lagrange_two_mode}).
The following equations of motion 
are obtained:
\beqn\label{simplest_EOM_coeffieints}
 & & \qquad \bcalH^{(2a \leftrightharpoons m)}(t) \C(t) = i \frac{\partial\C(t)}{\partial t}, \nonumber \\
 & &  {\mathcal H}_{p,p'}^{(2a \leftrightharpoons m)}(t) = 
\left<N-2p,p;t\left|\hat H^{(2a\leftrightharpoons m)} - i \frac{\partial}{\partial t} \right|N-2p',p';t\right>, \
\eeqn
where the vector $\C(t)$ collects the expansion coefficients $\{C_p(t)\}$.
Eq.~(\ref{simplest_EOM_coeffieints}) is a set of coupled first-order
differential equations with time-dependent coefficients,
and preserves the norm of an initially-normalized vector of coefficients $\C(0)$.
The time-dependent coefficients ${\mathcal H}_{p,p'}^{(2a \leftrightharpoons m)}(t)$,
being the matrix representation of $\hat H^{(2a\leftrightharpoons m)} - i \frac{\partial}{\partial t}$
in the subspace of configurations $\left\{\left|N-2p,p;t\right>\right\}$
and hence depending on the 
atomic $\phi_a(\r,t)$ and molecular $\psi_m(\r,t)$ orbitals,
are prescribed in the previous 
subsection \ref{two_mode_AM_energy}.

Next, 
we make use of the invariance of the multiconfigurational
wavefunction to unitary transformations (\ref{trans_appen_Psi_text}).
Explicitly,
the unitary transformation responsible for transforming
Eq.~(\ref{simplest_EOM_orbitals}) for the orbitals to 
Eq.~(\ref{simplest_EOM_orbitals_final}),
transforms equations of motion (\ref{simplest_EOM_coeffieints}) for the expansion coefficients to the final form:
\beqn\label{simplest_EOM_coeffieints_final}
 & & \qquad \H^{(2a \leftrightharpoons m)}(t) \C(t) = i \frac{\partial\C(t)}{\partial t}, \nonumber \\
 & &  H_{p,p'}^{(2a \leftrightharpoons m)}(t) = 
\left<N-2p,p;t\left|\hat H^{(2a\leftrightharpoons m)}\right|N-2p',p';t\right>. \
\eeqn
Equivalently,
Eq.~(\ref{simplest_EOM_coeffieints_final}) can be obtained from
Eq.~(\ref{simplest_EOM_coeffieints}) 
by substituting into
the latter 
the differential condition (\ref{MCTDHI_mix_conv_const_coherent}).

The coupled sets
of equations of motion 
for the atomic $\phi(\r,t)$ and molecular $\psi_m(\r,t)$ orbitals 
and expansion coefficients $\{C_p(t)\}$,  
Eqs.~(\ref{simplest_EOM_orbitals}) and (\ref{simplest_EOM_coeffieints})
or, respectively, Eqs.~(\ref{simplest_EOM_orbitals_final}) and (\ref{simplest_EOM_coeffieints_final})
constitute 
the {\it conversion mean field} theory 
(fully-variational two-mode approximation) 
for the interacting 
atomic--molecular 
system with 
conversion.

\subsection{The stationary self-consistent coherent mean field 
(time-independent fully-variational two-mode approximation)}\label{two_mode_AM_stationary}

The theory presented above is a time-dependent many-body theory.
It is certainly interesting to enquire what are the corresponding stationary many-body
states of the atomic--molecular Hamiltonian (\ref{ham_am_1}-\ref{ham_am_3})?
In other words,
what are the self-consistent solutions that minimize (extremize) the
expectation value $\left<\Psi\left|\hat H^{(2a\rightleftharpoons m)}\right|\Psi\right>$
for a given time-independent multiconfigurational ansatz 
$\left|\Psi\right> = \sum_{p=0}^{\left[N/2\right]} C_p \left|N-2p,p\right>$
assembled from time-independent atomic $\phi_a(\r)$ and molecular orbitals $\psi_m(\r)$?
To get this stationary self-consistent coherent mean field
(time-independent fully-variational two-mode approximation),
we resort to imaginary time-propagation
and set $t \to -it$ in the corresponding equations of motion.

Setting $t \to -it$ in 
(the left-hand side of)
either Eq.~(\ref{simplest_EOM_orbitals}) or (\ref{simplest_EOM_orbitals_final}),
the left-hand side decays to zero in time
and the equation becomes time-independent.
Then,
by multiplying the result, respectively, by
$\left<\hat N_a\right>$ and $\left<\hat N_m\right>$,
and translating back the projection
operators $\hat {\mathbf P}^{(a)}$ and $\hat {\mathbf P}^{(b)}$
to the corresponding Lagrange multipliers 
$\mu_a$ and $\mu_b$ (see appendix \ref{Lagrange_two_mode}),
we obtain the multiconfigurational
self-consistent (time-independent)
equations for the atomic and molecular orbitals:
\beqn\label{stationary_equations_orbitals}
& & \left[\left<\hat N_a\right> \hat h^{(a)} +
\lambda_a \left<\hat N_a(\hat N_a-1)\right> |\phi_a|^2 + 
 \lambda_{am} \left<\hat N_a \hat N_m\right> |\psi_m|^2 \right] \left|\phi_a\right> + \nonumber \\
 & & \qquad + \sqrt{2}\lambda_{con} \left<\hat b_a^\dag \hat b_a^\dag \hat c_m\right> 
 \phi_a^\ast \left|\psi_m\right> = \mu_a \left|\phi_a\right>, \nonumber \\
& & \left[\left<\hat N_m\right> \hat h^{(m)} +
\lambda_m \left<\hat N_m(\hat N_m-1)\right> |\psi_m|^2 +
 \lambda_{am} \left<\hat N_a \hat N_m\right> |\phi_a|^2 \right] \left|\psi_m\right> + \nonumber \\
 & & \qquad + \frac{\lambda_{con}}{\sqrt{2}} \left<\hat c_m^\dag \hat b_a \hat b_a\right> 
 \phi_a \left|\phi_a\right> = \mu_m \left|\psi_m\right>. \
\eeqn
Similarly,
restoring the Lagrange multiplier $\varepsilon(t)$ 
into either Eq.~(\ref{simplest_EOM_coeffieints}) 
or (\ref{simplest_EOM_coeffieints_final}),
see in this respect appendix \ref{Lagrange_two_mode},
and setting $t \to -it$ therein,
we obtain the stationary (self-consistent) eigenvalue equation
\beqn\label{stationary_equations_coeff}
 & & \qquad  \H^{(2a \leftrightharpoons m)} \C = \varepsilon \C, \nonumber \\ 
 & &  H_{p,p'}^{(2a \leftrightharpoons m)} = 
\left<N-2p,p\left|\hat H^{(2a\leftrightharpoons m)}\right|N-2p',p'\right> \
\eeqn
for the expansion coefficients.
We see that the (redundant) time-dependent Lagrange multiplier $\varepsilon(t)$
of the time-dependent theory has emerged as the
eigenenergy 
$\varepsilon = \left<\Psi\left|\hat H^{(2a\leftrightharpoons m)}\right|\Psi\right>$ 
of the stationary theory.

The theory ``distilled'' into Eqs.~(\ref{stationary_equations_orbitals},\ref{stationary_equations_coeff})
is a fully-variational stationary theory for the 
interacting atomic--molecular
system in presence of conversion,
where 
a single orbital is allowed for the atoms
and a single orbital to the molecules.
It is a system of 
coupled eigenvalue-like equations for the orbitals and eigenvalue equation for the coefficients,
thought 
non-linear and integrodifferential ones.

\section{The general multiconfigurational theory with atom--molecule conversion}\label{full_Hilbert}

In this section we develop a general many-body theory
for atom--molecule conversion,
by allowing 
the atoms and molecules to occupy more orbitals.
Section \ref{contact} builds the theory for the popular contact interaction,
whereas the case of generic non-contact interactions 
is presented in section \ref{non_contact}.

\subsection{Formulation for contact interactions}\label{contact}

\subsubsection{The multiconfigurational ansatz for the wavefunction}\label{contact_ansatz}

The multiconfigurational expansion mixes atomic--molecular states 
with different numbers of particles which are eigenfunctions of the 
particle-number operator $\hat N \left|\Psi(t)\right>=N\left|\Psi(t)\right>$:
\beqn\label{MCTDH_AM_Psi}
\left|\Psi(t)\right> &=& \sum_{p=0}^{[N/2]} \sum_{\vec{n}^p,\vec{m}^p}
C_{\vec{n}^p\vec{m}^p}(t) \left|\vec{n}^p,\vec{m}^p;t\right>, \nonumber \\ 
 \left|\vec{n}^p,\vec{m}^p;t\right> &\equiv& \frac{1}{\sqrt{n_1^p\cdots n_M^p! m_1^p!\cdots m_{M'}^p!}} \times \nonumber \\
 &\times& \left(\hat b_1^\dag(t)\right)^{n_1^p}\cdots\left(\hat b_M^\dag(t)\right)^{n_M^p} 
\left(\hat c_1^\dag(t)\right)^{m_1^p}\cdots\left(\hat c_{M'}^\dag(t)\right)^{m_{M'}^p}\left|vac\right>. \
\eeqn
We collect the individual occupations in the vectors 
$\vec{n}^p=(n_1^p,\ldots,n_M^p)$, $\vec{m}^p=(m_1,\ldots,m_{M'}^p)$.
The number of bosonic atoms $|\vec{n}^p| \equiv n_1^p+\ldots+n_M^p = N-2p$ 
and molecules $|\vec{m}^p| \equiv m_1^p+\ldots+m_{M'}^p = p$
of each configuration $\left|\vec{n}^p,\vec{m}^p;t\right>$ satisfies 
the particle-conservation law $|\vec{n}^p|+2|\vec{m}^p|=N$. 
Observe that the number of molecules $p$
serves as an index to the occupation numbers
$\vec{n}^p$ and $\vec{m}^p$.
This simply reflects the fact that,
for a given number of $N-2p$ atoms and $p$ molecules,
the possible occupation numbers which the configurations 
can assume
depend on $p$ itself.
The index $p$ together with the occupation numbers $\vec{n}^p$, $\vec{m}^p$
make a unique representation of each configuration.
The atomic and molecular number operators in the corresponding Hilbert space
boil down to $\hat N_a=\sum_{k=1}^M \hat b_k^\dag \hat b_k$ and 
$\hat N_m=\sum_{k'=1}^{M'} \hat c_{k'}^\dag \hat c_{k'}$, 
respectively.
The size of the resulting Hilbert space is given by
$\sum_{p=0}^{[N/2]} 
\begin{pmatrix}
 N-2p+M-1 \cr
 M-1
\end{pmatrix}
\begin{pmatrix}
 p+M'-1 \cr
 M'-1
\end{pmatrix}$,
i.e., by the sum of 
products of the sizes of the 
respective Hilbert subspaces for $N-2p$ bosonic atoms 
with $M$ orbitals
and $p$ bosonic molecules 
with $M'$ orbitals.

\subsubsection{Reduced density matrices for systems with particle conversion}\label{contact_reduced}

As part of the variational derivation we will need the expectation value of
$\hat H^{(2a\rightleftharpoons m)} - i\frac{\partial}{\partial t}$ 
with respect to $\Psi(t)$.
To this end,
it will be proved valuable to define and employ the
reduced density matrices of $\Psi(t)$.
We remind that 
L\"owdin has introduced the concept of
reduced density matrices 
for systems of a fixed number of 
particles (identical fermions) \cite{Lowdin}.
Nevertheless and although $\Psi(t)$ {\it is not} 
comprised of a fixed number of atoms 
or a fixed number of molecules,
it is possible to define 
the reduced density matrices
of a mixture of atoms and 
molecules {\it with conversion}. 

Having at hand the normalized many-body wavefunction $\Psi(t)$, 
the reduced one-body
density matrices of the atoms and molecules are defined by:  
\beqn\label{reduced_1B_2mix}
 & & \rho^{(a)}(\r_1|\r_2;t) = 
 \left<\hat{\mathbf \Psi}^\dag_a(\r_2)\hat{\mathbf \Psi}_a(\r_1)\right> =
 \sum^M_{k,q=1} \rho^{(a)}_{kq}(t) \phi^\ast_k(\r_2,t)\phi_q(\r_1,t), \nonumber \\
 & & \rho^{(m)}(\r_1|\r_2;t) = 
 \left<\hat{\mathbf \Psi}^\dag_m(\r_2)\hat{\mathbf \Psi}_m(\r_1)\right> =
 \sum^{M'}_{k',q'=1} \rho^{(m)}_{k'q'}(t) \psi^\ast_{k'}(\r_2,t)\psi_{q'}(\r_1,t), \
\eeqn
where the matrix elements 
$\rho_{kq}^{(a)}(t) = \left<\hat b_k^\dag \hat b_q\right>$ and 
$\rho_{k'q'}^{(m)}(t) = \left<\hat c_{k'}^\dag \hat c_{q'}\right>$
are prescribed in appendix \ref{matrix_Appen}.
We collect these matrix elements as 
$\brho^{(a)}(t)=\left\{\rho_{kq}^{(a)}(t)\right\}$ and 
$\brho^{(m)}(t)=\left\{\rho_{k'q'}^{(m)}(t)\right\}$.
Similarly,
the reduced two-body density matrices of the atoms and molecules are defined by:
\beqn\label{reduced_2B_2mix}
 & & \rho^{(a)}(\r_1,\r_2|\r_3,\r_4;t) = 
 \left<\hat{\mathbf \Psi}^\dag_a(\r_3)\hat{\mathbf \Psi}^\dag_a(\r_4)\hat{\mathbf \Psi}_a(\r_2)
\hat{\mathbf \Psi}_a(\r_1)\right> = \nonumber \\
 & & \qquad = 
\sum^M_{k,s,l,q=1} \rho^{(a)}_{kslq}(t) \phi^\ast_k(\r_3,t)\phi^\ast_s(\r_4,t)\phi_l(\r_2,t)\phi_q(\r_1,t), \nonumber \\
 & & \rho^{(m)}(\r_1,\r_2|\r_3,\r_4;t) = 
 \left<\hat{\mathbf \Psi}^\dag_m(\r_3)\hat{\mathbf \Psi}^\dag_m(\r_4)\hat{\mathbf \Psi}_m(\r_2)
\hat{\mathbf \Psi}_m(\r_1)\right> = \nonumber \\
 & & \qquad = \sum^{M'}_{k',s',l',q'=1} \rho^{(m)}_{k's'l'q'}(t) 
\psi^\ast_{k'}(\r_3,t)\psi^\ast_{s'}(\r_4,t)\psi_{l'}(\r_2,t)\psi_{q'}(\r_1,t), \nonumber \\
 & & \rho^{(am)}(\r_1,\r_2|\r_3,\r_4;t) = 
 \left<\hat{\mathbf \Psi}^\dag_a(\r_3)\hat{\mathbf \Psi}_a(\r_1)\hat{\mathbf \Psi}^\dag_m(\r_4)
\hat{\mathbf \Psi}_m(\r_2)\right> = \nonumber \\
 & & \qquad = \sum^{M}_{k,q=1} \sum^{M'}_{k',q'=1} \rho^{(am)}_{kk'qq'}(t) 
\phi^\ast_{k}(\r_3,t)\phi_{q}(\r_1,t)\psi^\ast_{k'}(\r_4,t)\psi_{q'}(\r_2,t), \
\eeqn
where the matrix elements
$\rho_{kslq}^{(a)}(t) = \left<\hat b_k^\dag \hat b_s^\dag \hat b_l \hat b_q\right>$,
$\rho_{k's'l'q'}^{(m)}(t) = \left<\hat c_{k'}^\dag \hat c_{s'}^\dag \hat c_{l'} \hat c_{q'}\right>$,
and
$\rho_{kk'qq'}^{(am)}(t) = \left<\hat b_k^\dag \hat b_q \hat c_{k'}^\dag \hat c_{q'}\right>$
are prescribed in appendix \ref{matrix_Appen}.
Because the reduced density matrices 
(\ref{reduced_1B_2mix}) and (\ref{reduced_2B_2mix})
directly only couple 
configurations with the same number of atoms and molecules,
we will refer to them 
as {\it particle-conserving} 
reduced density matrices.
In this context, 
$\rho^{(am)}(\r_1,\r_2|\r_3,\r_4;t)$
is the lowest-order 
{\it inter-species} 
particle-conserving 
reduced density matrix.

From the above discussion it is anticipated that,
due to the conversion term (\ref{ham_am_3}) in the Hamiltonian, 
another kind of reduced density matrices appear in the theory.
Specifically,
we define the 
{\it particle non-conserving} 
reduced density matrices as follows:
\beqn\label{reduced_conver}
 & &  \!\!\!\!\!\! \rho^{(2a\rightharpoonup m)}(\r_1,\r_2|\r_3;t) = 
 \left<\hat{\mathbf \Psi}^\dag_m(\r_3)\hat{\mathbf \Psi}_a(\r_2)\hat{\mathbf \Psi}_a(\r_1)\right> =  
  \sum^{M'}_{k'=1} \sum^{M}_{k,q=1} \rho^{(2a\rightharpoonup m)}_{k'kq}(t) 
 \psi^\ast_{k'}(\r_3,t)\phi_{k}(\r_2,t)\phi_{q}(\r_1,t), \nonumber \\
 & &  \!\!\!\!\!\! \rho^{(m\rightharpoondown 2a)}(\r_3|\r_2,\r_1;t) =
 \left<\hat{\mathbf \Psi}^\dag_a(\r_1)\hat{\mathbf \Psi}^\dag_a(\r_2)\hat{\mathbf \Psi}_m(\r_3)\right> = 
 \sum^{M'}_{k'=1} \sum^{M}_{k,q=1} \rho^{(m\rightharpoondown 2a)}_{qkk'}(t)
 \phi^\ast_q(\r_1,t) \phi^\ast_k(\r_2,t) \psi_{k'}(\r_3,t), \nonumber \\
 & & \rho^{(m\rightharpoondown 2a)}(\r_3|\r_2,\r_1;t) = 
\left\{\rho^{(2a\rightharpoonup m)}(\r_1,\r_2|\r_3;t)\right\}^\ast, \qquad 
  \rho^{(m\rightharpoondown 2a)}_{qkk'}(t) = \left\{\rho^{(2a\rightharpoonup m)}_{k'kq}(t)\right\}^\ast.
\eeqn
The matrix elements 
$\rho^{(2a\rightharpoonup m)}_{k'kq} = \left<\hat c_{k'}^\dag \hat b_k \hat b_q\right>$
are given in appendix \ref{matrix_Appen}.

\subsubsection{The functional action $S$ and its evaluation}\label{contact_func}

We start from the functional action of the time-dependent Schr\"odinger 
equation which in the general multiconfigurational 
case takes on the form:
\beqn\label{action_functional_AM_full}
& & 
S\left[\{C_{\vec{n}^p\vec{m}^p}(t)\},\left\{\phi_k(\r,t)\right\},\left\{\psi_{k'}(\r,t)\right\}\right] =  
 \int dt \Bigg\{\left<\Psi(t)\left| \hat H^{(2a\leftrightharpoons m)} 
- i\frac{\partial}{\partial t} \right|\Psi(t)\right> \nonumber \\
& & \ \ - 
\sum_{k,j=1}^{M} \mu^{(a)}_{kj}(t)\left[\left<\phi_k(\r,t)|\phi_j(\r,t)\right> - 
\delta_{kj} \right]  - \sum_{k',j'=1}^{M'} \mu^{(m)}_{k'j'}(t) \left[\left<\psi_{k'}(\r,t)|\psi_{j'}(\r,t)\right> - 
\delta_{k'j'} \right] \nonumber \\
& & \ \ - 
\varepsilon(t)\left[ \sum_{p=0}^{[N/2]} \sum_{\vec{n}^p,\vec{m}^p} 
 \left|C_{\vec{n}^p\vec{m}^p}(t)\right|^2 - 1\right]\Bigg\}. \ 
\eeqn
The time-dependent Lagrange multiplies $\{\mu^{(a)}_{kj}(t)\}$, $\{\mu^{(m)}_{k'j'}(t)\}$ and $\varepsilon(t)$
are introduced to ensure orthonormalization 
of the atomic $\left\{\phi_k(\r,t)\right\}$
and molecular $\left\{\psi_{k'}(\r,t)\right\}$ 
orbital sets 
and normalization 
of the expansion coefficients $\{C_{\vec{n}^p\vec{m}^p}(t)\}$.

To derive the 
equations of motion for the atomic--molecular multiconfigurational 
wavefunction (\ref{MCTDH_AM_Psi}),
the expectation value of 
$\hat H^{(2a\rightleftharpoons m)} - i\frac{\partial}{\partial t}$ with respect to $\Psi(t)$ is needed,
where $\hat H^{(2a\rightleftharpoons m)}$ is given in Eq.~(\ref{ham_contact_inter}).
The expectation value of 
$\hat H^{(2a\rightleftharpoons m)} - i\frac{\partial}{\partial t}$ 
is expressed by two equivalent forms,
as done in section \ref{two_mode_AM_energy}.
The first form, 
where 
the dependence of 
Eq.~(\ref{action_functional_AM_full}) on the atomic and molecular orbitals 
is explicit, 
reads:
\beqn\label{expectation_ham_a}
& & 
\left<\Psi(t)\left|\hat H^{(2a\rightleftharpoons m)} - i\frac{\partial}{\partial t}\right|\Psi(t)\right> = 
\sum_{k,q=1}^M \rho^{(a)}_{kq} \left<\phi_k\left|h^{(a)} - i\frac{\partial}{\partial t}\right|\phi_q\right> + \nonumber \\
& & +
\frac{\lambda_a}{2}\sum_{k,s,l,q=1}^M \rho^{(a)}_{kslq} \left<\phi_k\phi_s|\phi_q\phi_l\right> + 
\sum_{k',q'=1}^{M'} \rho^{(m)}_{k'q'} 
\left<\psi_{k'}\left|h^{(m)} - i\frac{\partial}{\partial t}\right|\psi_{q'}\right> +  \\
& & + \frac{\lambda_m}{2}\sum_{k',s',l',q'=1}^{M'} \rho^{(m)}_{k's'l'q'} \left<\phi_{k'}\phi_{s'}|\phi_{q'}\phi_{l'}\right> + 
\lambda_{am}\sum_{k,q=1}^M \sum_{k',q'=1}^{M'} \rho^{(am)}_{kk'qq'} \left<\phi_k\psi_{k'}|\phi_q\psi_{q'}\right> + \nonumber \\
& & + \frac{\lambda_{con}}{\sqrt{2}} 
\sum_{k,q=1}^{M} \sum_{k'=1}^{M'} 
\left[
\rho_{k'kq}^{(2a\rightharpoonup m)}
\left<\psi_{k'}\left|\right.\phi_k \phi_q\right> +
\rho_{qkk'}^{(m\rightharpoondown 2a)}
\left<\phi_q \phi_k\left|\right.\psi_{k'}\right> 
\right] 
- i  \sum_{p=0}^{\left[N/2\right]} \sum_{\vec{n}^p,\vec{m}^p}
 C_{\vec{n}^p\vec{m}^p}^\ast \frac{\partial C_{\vec{n}^p\vec{m}^p}}{\partial t}. \nonumber \
\eeqn 
We see in (\ref{expectation_ham_a}) the appearance of
the particle-conserving and particle non-conserving reduced density matrices 
introduced in the previous subsection 
\ref{contact_reduced}.
Eq.~(\ref{expectation_ham_a})
is to be used to derive the equations of motion of
$\left\{\phi_k(\r,t)\right\}$ and $\left\{\psi_{k'}(\r,t)\right\}$.

The second form of the expectation value of $\hat H^{(2a\rightleftharpoons m)} - i\frac{\partial}{\partial t}$
in the functional action (\ref{action_functional_AM_full}),
\beqn\label{expectation_ham_b} 
& & \left<\Psi(t)\left|\hat H^{(2a\rightleftharpoons m)} - i\frac{\partial}{\partial t}\right|\Psi(t)\right> 
 = \sum_{p=0}^{\left[N/2\right]} \sum_{\vec{n}^p,\vec{m}^p}
C_{\vec{n}^p\vec{m}^p}^\ast \times \nonumber \\
& & \times \Bigg[\sum_{p'=0}^{\left[N/2\right]} \sum_{\vec{n}'^{p'},\vec{m}'^{p'}}
\left<\vec{n}^p,\vec{m}^p;t\left|\hat H^{(2a\leftrightharpoons m)} - 
i\frac{\partial}{\partial t}\right|\vec{n}'^{p'},\vec{m}'^{p'};t\right> 
C_{\vec{n}'^{p'},\vec{m}'^{p'}} - i \frac{\partial C_{\vec{n}^p\vec{m}^p}}{\partial t}\Bigg], \
\eeqn
displays its explicit 
dependence on the expansion coefficients,
and therefore will be employed to derive 
the equations of motion of $\{C_{\vec{n}^p\vec{m}^p}(t)\}$.
Finally, 
it is deductive to
compare 
the structure 
of Eqs.~(\ref{Hamiltonian_matrix_element_AM_a},\ref{Hamiltonian_matrix_element_AM_b}) in the 
conversion 
mean field 
(fully-variational two-mode) problem
to that of 
Eqs.~(\ref{expectation_ham_a},\ref{expectation_ham_b})
of the general problem.

\subsubsection{The equations of motion for $\Psi(t)$}\label{contact_equation}

Collecting the above ingredients,
we are ready to perform 
the variation of the functional action
$S\left[\{C_{\vec{n}^p\vec{m}^p}(t)\},\left\{\phi_k(\r,t)\right\},\left\{\psi_{k'}(\r,t)\right\}\right]$
and arrive at the equations
of motion of $\Psi(t)$.
Equating the variation of the action functional 
(\ref{action_functional_AM_full},\ref{expectation_ham_a})
with respect to the orbitals to zero
and eliminating the Lagrange multipliers $\{\mu_{kj}^{(a)}(t)\}$, $\{\mu_{k'j'}^{(m)}(t)\}$
(see appendix \ref{appen_C}),
we obtain 
the following result, $j=1,\ldots,M$, $j'=1,\ldots,M'$:
\beqn\label{MCTDH_conver_orb_eqs_P}
& & \!\!\!\!\!\!\!\! \hat{\mathbf P}^{(a)} i\left|\dot\phi_j\right> =
   \hat{\mathbf P}^{(a)} \Bigg[\hat h^{(a)} \left|\phi_j\right> + \nonumber \\
& & + \sum^M_{k=1} \left\{\brho^{(a)}(t)\right\}^{-1}_{jk} \times  \sum^M_{q=1}
 \Bigg( \left\{\rho_2(\phi^2,\psi^2)\right\}^{(a)}_{kq} \left|\phi_q\right> 
 + \sqrt{2} \lambda_{con} \sum_{k'=1}^{M'} 
\rho^{(m\rightharpoondown 2a)}_{qkk'} \phi_q^\ast \left|\psi_{k'}\right> \Bigg) \Bigg], \nonumber \\
& & \!\!\!\!\!\!\!\! \hat{\mathbf P}^{(m)} i\left|\dot\psi_{j'}\right> =
  \hat{\mathbf P}^{(m)} \Bigg[\hat h^{(m)} \left|\psi_{j'}\right> + \nonumber \\
& & + \sum^{M'}_{k'=1} \left\{\brho^{(m)}(t)\right\}^{-1}_{j'k'} \times
\Bigg( \sum^{M'}_{q'=1} \left\{\rho_2(\psi^2,\phi^2)\right\}^{(m)}_{k'q'}
\left|\psi_{q'}\right> 
 + \frac{\lambda_{con}}{\sqrt{2}} \sum_{k,q=1}^{M} 
\rho^{(2a\rightharpoonup m)}_{kk'q} \phi_k \left|\phi_{q}\right> \Bigg) \Bigg], \
\eeqn
where terms 
with products of 
reduced two-body density matrices 
times orbital pairs 
are collected together
and denoted 
for brevity as
\beqn\label{rho_phi}
 & & \left\{\rho_2(\phi^2,\psi^2)\right\}^{(a)}_{kq} \equiv 
\lambda_a \sum^M_{s,l=1} \rho^{(a)}_{kslq} (\phi_s^\ast \phi_l) + 
\lambda_{am} \sum_{k',q'=1}^{M'} \rho^{(am)}_{kk'qq'} (\psi_{k'}^\ast\psi_{q'}), \nonumber \\
 & & \left\{\rho_2(\psi^2,\phi^2)\right\}^{(m)}_{k'q'} \equiv 
\lambda_m \sum^{M'}_{s',l'=1} \rho^{(m)}_{k's'l'q'} (\psi_{s'}^\ast \psi_{l'}) + 
\lambda_{am} \sum_{k,q=1}^{M} \rho^{(am)}_{kk'qq'} (\phi_{k}^\ast\phi_{q}), \
\eeqn
and
\beq\label{projection_conver}
\hat {\mathbf P}^{(a)} = 1 - \sum_{u=1}^{M}\left|\phi_u\left>\right<\phi_u\right|, \qquad
\hat {\mathbf P}^{(m)} = 1 - \sum_{u'=1}^{M'}\left|\psi_{u'}\left>\right<\psi_{u'}\right|,
\eeq
are projection operators.
When acting on one-body functions in the atomic and molecular spaces,
$\hat {\mathbf P}^{(a)}$ and $\hat {\mathbf P}^{(m)}$
project them onto the subspaces
orthogonal to those spanned by the orbitals
$\left\{\phi_k(\r,t)\right\}$ and $\left\{\psi_{k'}(\r,t)\right\}$,
respectively.
These projection operators emerge when one eliminates the Lagrange multipliers
$\{\mu_{kj}^{(a)}(t)\}$ and $\{\mu_{k'j'}^{(m)}(t)\}$
from the equations of motion, 
see appendix \ref{appen_C}.
To remind,
we use the shorthand 
notation $\dot\phi_j \equiv \frac{\partial\phi_j}{\partial t}$,
$\dot\psi_{j'} \equiv \frac{\partial\psi_{j'}}{\partial t}$
in the equations of motion for the orbitals.

The appearance of the 
projection operators on both
the left- and right-hand sides of Eq.~(\ref{MCTDH_conver_orb_eqs_P})
makes the system (\ref{MCTDH_conver_orb_eqs_P})
a cumbersome system of integrodifferential non-linear equations.
This situation can be simplified 
by generalizing the treatment of section \ref{two_mode_AM_EOM},
namely,
exploiting the
invariance properties
of the many-particle wavefunction $\Psi(t)$.
Specifically, 
we perform independent unitary transformations on the
atomic $\left\{\phi_k(\r,t)\right\} \to \left\{\overline{\phi}_k(\r,t)\right\}$ and molecular 
$\left\{\psi_{k'}(\r,t)\right\} \to \left\{\overline{\psi}_{k'}(\r,t)\right\}$ orbitals,
which results in transformed configurations 
$\left|\vec{n}^p,\vec{m}^p;t\right> \to \overline{\left|\vec{n}^p,\vec{m}^p;t\right>}$.
Then, 
we can compensate for the transformations of the orbital sets 
by the ``reverse'' transformation of the expansion coefficients 
$\{C_{\vec{n}^p\vec{m}^p}(t)\} \to \{\overline{C}_{\vec{n}^p\vec{m}^p}(t)\}$.
We represent this invariance by the following equality:
\beq\label{invariance_rotations}
 \left|\Psi(t)\right> = 
\sum_{p=0}^{[N/2]} \sum_{\vec{n}^p,\vec{m}^p}
C_{\vec{n}^p\vec{m}^p}(t) \left|\vec{n}^p,\vec{m}^p;t\right> = 
\sum_{p=0}^{[N/2]} \sum_{\vec{n}^p,\vec{m}^p}
\overline{C}_{\vec{n}^p\vec{m}^p}(t) \overline{\left|\vec{n}^p,\vec{m}^p;t\right>}.
\eeq
The transformations of the 
orbital sets 
and expansion coefficients do not change the size
of the Hilbert space,
or couple systems with different numbers of atoms, molecules.
We remark that 
transformations which inter-mix atomic and molecular 
orbitals are not required for our needs. 
To represent these properties, 
the same occupation numbers
$\vec{n}^p$, $\vec{m}^p$ and summation 
index $p$ of the number of molecules 
are used for both 
multiconfigurational expansions of $\Psi(t)$ 
in Eq.~(\ref{invariance_rotations}).

We can now make use of the invariance (\ref{invariance_rotations})
to simplify the equations of motion (\ref{MCTDH_conver_orb_eqs_P}),
without introducing further constraints into the equations of motion.
We utilize 
a specific unitary transformation of the many-particle 
wavefunction that 
eliminates the projection operators
acting on the time-derivatives (left-hand sides)
in Eq.~(\ref{MCTDH_conver_orb_eqs_P});
see appendix \ref{appen_C} for more details.
The final result 
for the equations of motion for the
atomic $\left\{\phi_k(\r,t)\right\}$ and molecular $\left\{\psi_{k'}(\r,t)\right\}$ 
orbitals thus takes on the form,
$j=1,\ldots,M$, $j'=1,\ldots,M'$:
\beqn\label{MCTDH_conver_orb_eqs}
& & \!\!\!\!\!\!\!\! i\left|\dot\phi_j\right> =
   \hat{\mathbf P}^{(a)} \Bigg[\hat h^{(a)} \left|\phi_j\right> + \nonumber \\
& & + \sum^M_{k=1} \left\{\brho^{(a)}(t)\right\}^{-1}_{jk} \times  \sum^M_{q=1}
 \Bigg( \left\{\rho_2(\phi^2,\psi^2)\right\}^{(a)}_{kq} \left|\phi_q\right> 
 + \sqrt{2} \lambda_{con} \sum_{k'=1}^{M'} 
\rho^{(m\rightharpoondown 2a)}_{qkk'} \phi_q^\ast \left|\psi_{k'}\right> \Bigg) \Bigg], \nonumber \\
& & \!\!\!\!\!\!\!\! i\left|\dot\psi_{j'}\right> =
  \hat{\mathbf P}^{(m)} \Bigg[\hat h^{(m)} \left|\psi_{j'}\right> + \nonumber \\
& & + \sum^{M'}_{k'=1} \left\{\brho^{(m)}(t)\right\}^{-1}_{j'k'} \times
\Bigg( \sum^{M'}_{q'=1} \left\{\rho_2(\psi^2,\phi^2)\right\}^{(m)}_{k'q'}
\left|\psi_{q'}\right> 
 + \frac{\lambda_{con}}{\sqrt{2}} \sum_{k,q=1}^{M} 
\rho^{(2a\rightharpoonup m)}_{kk'q} \phi_k \left|\phi_{q}\right> \Bigg) \Bigg], \
\eeqn
with the projection operators 
$\hat {\mathbf P}^{(a)}$ and 
$\hat {\mathbf P}^{(m)}$ 
appearing
now on the right-hand sides only.

Now,
taking the respective scalar products of (\ref{MCTDH_conver_orb_eqs})
with $\left\{\left<\phi_k\right|\right\}$ and $\left\{\left<\psi_{k'}\right|\right\}$,
we obtain the 
following differential conditions: 
\beq\label{MCTDHI_mix_conv_const}
\left<\phi_k\left|\right.\dot\phi_q\right> = 0, \ \ k,q=1,\ldots,M, \qquad
\left<\psi_{k'}\left|\right.\dot\psi_{q'}\right> = 0, \ \ k',q'=1,\ldots,M'.
\eeq
It is instructive to mention
that these differential condition
have been introduced originally by the 
MCTDH developers \cite{CPL,JCP},
and used thereafter
in particle-conserving 
multiconfigurational theories for identical particles and mixtures
\cite{MCTDHF1,MCTDHF2,MCTDHF3,MCTDHB1,MCTDHB2,Unified_paper,MIX}.

The differential conditions (\ref{MCTDHI_mix_conv_const})
ensure that initially-orthonormalized
orbital sets $\left\{\phi_k(\r,t)\right\}$, $\left\{\psi_{k'}(\r,t)\right\}$ 
remain orthonormalized at all times.
The meaning of the unitary transformation
carrying equations of motion (\ref{MCTDH_conver_orb_eqs_P})
to (\ref{MCTDH_conver_orb_eqs}) can now be seen.
This unitary 
transformation takes orthonormal time-dependent orbitals,
$\left<\phi_k\left|\right.\phi_q\right> = \delta_{kq}$ and $\left<\psi_{k'}\left|\right.\psi_{q'}\right> = \delta_{k'q'}$
[$\delta_{kq}$, $\delta_{k'q'}$ is the Dirac delta-function],
which therefore satisfy the general relations
$\frac{\partial \left<\phi_k\left|\right.\phi_q\right>}{\partial t} = 
\left<\dot\phi_k\left|\right.\phi_q\right> + \left<\phi_k\left|\right.\dot\phi_q\right> = 0$
and
$\frac{\partial \left<\psi_{k'}\left|\right.\psi_{q'}\right>}{\partial t} = 
\left<\dot\psi_{k'}\left|\right.\psi_{q'}\right> + \left<\psi_{k'}\left|\right.\dot\psi_{q'}\right> = 0$,
and transforms them to
time-dependent orbitals
satisfying the specific differential conditions 
(\ref{MCTDHI_mix_conv_const}).

Moving to the equations of motion for the coefficients 
$\{C_{\vec{n}^p\vec{m}^p}(t)\}$,
we equate the variation of the
functional action
(\ref{action_functional_AM_full},\ref{expectation_ham_b})
with respect to the expansion coefficients to zero.
Eliminating the Lagrange multiplier $\varepsilon(t)$ 
by a respective 
phase transformation of the coefficients,
we arrive at the form:
\beqn\label{MCTDH_conver_coeff_eqs_P}
& & \qquad {\bcalH}^{(2a \leftrightharpoons m)}(t)\C(t) = i\frac{\partial \C(t)}{\partial t}, \nonumber \\
& &  {\cal H}^{(2a \leftrightharpoons m)}_{\vec{n}^p\vec{m}^p,\vec{n}'^{p'}\vec{m}'^{p'}}(t) =
\left<\vec{n}^p,\vec{m}^p;t\left|\hat H^{(2a \leftrightharpoons m)} - i\frac{\partial}{\partial t}
\right|\vec{n}'^{p'}\vec{m}'^{p'};t\right>. \
\eeqn
Eq.~(\ref{MCTDH_conver_coeff_eqs_P})
has exactly the same form as 
Eq.~(\ref{simplest_EOM_coeffieints}) in the specific theory of 
section \ref{two_mode_AM_EOM},
and constitutes a set of coupled first-order
differential equations with time-dependent coefficients
that preserve 
the norm of an initially-normalized vector of coefficients $\C(0)$.
The matrix elements of $\hat H^{(2a\rightleftharpoons m)} - i\frac{\partial}{\partial t}$ 
with respect to two general configurations,
${\cal H}^{(2a \leftrightharpoons m)}_{\vec{n}^p\vec{m}^p,\vec{n}'^{p'}\vec{m}'^{p'}}(t)$,
are prescribed in appendix \ref{matrix_Appen}.

Now, 
to arrive at the final form of the equations of motion for the coefficients,
we make use of
the invariance of $\Psi(t)$ 
to unitary transformations (\ref{invariance_rotations}).
Specifically,
the unitary transformation 
carrying
Eq.~(\ref{MCTDH_conver_orb_eqs_P}) for the orbitals to Eq.~(\ref{MCTDH_conver_orb_eqs}),
casts 
Eq.~(\ref{MCTDH_conver_coeff_eqs_P}) for the expansion coefficients 
into the final result:
\beqn\label{MCTDH_conver_coeff_eqs}
& & \qquad {\mathbf H}^{(2a \leftrightharpoons m)}(t)\C(t) = i\frac{\partial \C(t)}{\partial t}, \nonumber \\
& &  H^{(2a \leftrightharpoons m)}_{\vec{n}^p\vec{m}^p,\vec{n}'^{p'}\vec{m}'^{p'}}(t) =
\left<\vec{n}^p,\vec{m}^p;t\left|\hat H^{(2a \leftrightharpoons m)} \right|\vec{n}'^{p'}\vec{m}'^{p'};t\right>. \
\eeqn
Eq.~(\ref{MCTDH_conver_coeff_eqs}) has exactly the same form  
as Eq.~(\ref{simplest_EOM_coeffieints_final}) in the 
specific case of section \ref{two_mode_AM_EOM}.
Of course,
there are much more expansion coefficients in the general case.
Finally and equivalently,
we note that 
the result (\ref{MCTDH_conver_coeff_eqs})
can be obtained from 
Eq.~(\ref{MCTDH_conver_coeff_eqs_P}) when
the differential 
condition (\ref{MCTDHI_mix_conv_const})
is substituted 
into the latter.

Let us pause for a moment and summarize.
We have started from the functional action
(\ref{action_functional_AM_full})
and arrived at the equations of motion for $\Psi(t)$.
Eq.~(\ref{MCTDH_conver_orb_eqs_P}) for the orbitals 
$\left\{\phi_k(\r,t)\right\}$, $\left\{\psi_{k'}(\r,t)\right\}$
and Eq.~(\ref{MCTDH_conver_coeff_eqs_P}) for the expansion
coefficients $\{C_{\vec{n}^p\vec{m}^p}(t)\}$,
or, respectively,
Eqs.~(\ref{MCTDH_conver_orb_eqs}) and 
(\ref{MCTDH_conver_coeff_eqs})
constitute the multiconfigurational time-dependent Hartree 
theory for systems with particle conversion;
here specifically the theory for 
bosonic atoms and 
bosonic molecules 
with conversion.

\subsubsection{The stationary self-consistent 
general multiconfigurational theory 
with conversion}\label{contact_equation_stationary}

The theory presented above is 
a time-dependent many-body theory and, 
as done in the previous section \ref{two_mode_AM},
it is relevant to put forward the corresponding
stationary general theory.
Consider the multiconfigurational expansion
$\left|\Psi\right> = \sum_{p=0}^{[N/2]} \sum_{\vec{n}^p,\vec{m}^p}
C_{\vec{n}^p\vec{m}^p}\left|\vec{n}^p,\vec{m}^p\right>$,
where the expansion coefficients 
$\{C_{\vec{n}^p\vec{m}^p}\}$
and orbitals 
$\left\{\phi_k(\r)\right\}$, $\left\{\psi_{k'}(\r)\right\}$
assembling the configurations 
$\left\{\left|\vec{n}^p,\vec{m}^p\right>\right\}$
are time-independent quantities.
What are then the self-consistent solutions
that minimize (extremize) the expectation value
$\left<\Psi\left|\hat H^{(2a\leftrightharpoons m)}\right|\Psi\right>$?

The working equations of the stationary 
theory can be obtained
by resorting to imaginary time-propagation and 
setting $t \to -it$ in either 
equations of motion (\ref{MCTDH_conver_orb_eqs_P}) and (\ref{MCTDH_conver_coeff_eqs_P})
or equations of motion (\ref{MCTDH_conver_orb_eqs}) and (\ref{MCTDH_conver_coeff_eqs}),
after the time-dependent Lagrange multiplier $\varepsilon(t)$ has been reinstated.
Then,
by translating the projection operators 
$\hat{\mathbf P}^{(a)}$, $\hat{\mathbf P}^{(m)}$
to the respective Lagrange multipliers 
$\{\mu^{(a)}_{kj}\}$, $\{\mu^{(m)}_{k'j'}\}$,
the resulting working equations take on the form:
\beqn\label{MCTDH_conver_stationary_orb}
& & \!\!\!\!\!\!\!\!\!\! \sum_{q=1}^M \Bigg\{ \Bigg[ \rho^{(a)}_{kq}\hat h^{(a)} +
 \left\{\rho_2(\phi^2,\psi^2)\right\}^{(a)}_{kq}
\Bigg] \left|\phi_q\right> + \sqrt{2} \lambda_{con} \sum_{k'=1}^{M'} 
\rho^{(m\rightharpoondown 2a)}_{qkk'} \phi_q^\ast \left|\psi_{k'}\right> \Bigg\}
= \sum_{j=1}^M \mu^{(a)}_{kj} \left|\phi_j\right>, \nonumber \\ & & \qquad k=1,\ldots,M, \nonumber \\
& & \!\!\!\!\!\!\!\!\!\! \sum_{q'=1}^{M'} \Bigg[ \rho^{(m)}_{k'q'}\hat h^{(m)} +
 \left\{\rho_2(\psi^2,\phi^2)\right\}^{(m)}_{k'q'}
\Bigg] \left|\psi_{q'}\right> + \frac{\lambda_{con}}{\sqrt{2}} \sum_{k,q=1}^{M} 
\rho^{(2a\rightharpoonup m)}_{k'kq} \phi_k \left|\phi_{q}\right>
= \sum_{j'=1}^{M'} \mu^{(m)}_{k'j'} \left|\psi_{j'}\right>, \nonumber \\ & & \qquad k'=1,\ldots,M' \
\eeqn
for the orbitals, 
and 
\beqn\label{MCTDH_conver_stationary_coeff}
 & & \qquad {\mathbf H}^{(2a\leftrightharpoons m)} \C = \varepsilon \C, 
\nonumber \\
 & &  H^{(2a \leftrightharpoons m)}_{\vec{n}^p\vec{m}^p,\vec{n}'^{p'}\vec{m}'^{p'}} =
\left<\vec{n}^p,\vec{m}^p\left|\hat H^{(2a \leftrightharpoons m)} \right|\vec{n}'^{p'},\vec{m}'^{p'}\right> \
\eeqn
for the expansion coefficients.
As seen in section \ref{two_mode_AM},
the time-dependent Lagrange multiplier $\varepsilon(t)$
emerges 
in the time-independent theory
as the eigenenergy of the coupled atom--molecule system with
$\varepsilon=\left<\Psi\left|\hat H^{(2a\leftrightharpoons m)}\right|\Psi\right>$.

The stationary equations for the 
orbitals (\ref{MCTDH_conver_stationary_orb})
can be further 
simplified.
The stationary wavefunction $\Psi$,
as its time-dependent counterpart,
is invariant to independent unitary
transformations of the orbital sets 
$\left\{\phi_k(\r)\right\}$, $\left\{\psi_{k'}(\r)\right\}$
and the ``inverse'' transformation of the expansion coefficients $\{C_{\vec{n}^p\vec{m}^p}\}$.
We can use the unitary matrices
which diagonalize 
the matrices of Lagrange multipliers
$\{\mu^{(a)}_{kj}\}$, $\{\mu^{(m)}_{k'j'}\}$. 
We note that the
matrices of Lagrange multipliers
are Hermitian matrices for stationary states.
As a result of this transformation,
we obtain a set of coupled 
equations for the orbitals that look just as Eq.~(\ref{MCTDH_conver_stationary_orb}),
except for the right-hand sides being diagonal:
\beqn\label{diagonal_form}
& & \!\!\!\!\!\!\!\!\!\! \sum_{q=1}^M \Bigg\{ \Bigg[ \rho^{(a)}_{kq}\hat h^{(a)} +
\left\{\rho_2(\phi^2,\psi^2)\right\}^{(a)}_{kq} \Bigg] \left|\phi_q\right> 
+ \sqrt{2} \lambda_{con} \sum_{k'=1}^{M'} 
\rho^{(m\rightharpoondown 2a)}_{qkk'} \phi_q^\ast \left|\psi_{k'}\right> \Bigg\}
= \mu^{(a)}_{k} \left|\phi_k\right>, \nonumber \\ & & \qquad k=1,\ldots,M, \nonumber \\
& & \!\!\!\!\!\!\!\!\!\! \sum_{q'=1}^{M'} \Bigg[ \rho^{(m)}_{k'q'}\hat h^{(m)} +
\left\{\rho_2(\psi^2,\phi^2)\right\}^{(m)}_{k'q'}
\Bigg] \left|\psi_{q'}\right> 
+ \frac{\lambda_{con}}{\sqrt{2}} \sum_{k,q=1}^{M} 
\rho^{(2a\rightharpoonup m)}_{k'kq} \phi_k \left|\phi_{q}\right>
= \mu^{(m)}_{k'}\left|\psi_{k'}\right>, \nonumber \\ & & \qquad k'=1,\ldots,M'. \
\eeqn
The form of the equation for the 
expansion coefficients (\ref{MCTDH_conver_stationary_coeff}) does not change.
Thus, 
the final result for the stationary theory,
Eqs.~(\ref{MCTDH_conver_stationary_coeff},\ref{diagonal_form}),
is 
a coupled system of integrodifferential,
non-linear equations 
constituting eigenvalue-like equations 
for the orbitals and eigenvalue equation for
the expansion coefficients; 
compare to section \ref{two_mode_AM_stationary}.
Generally,
the transformation of the matrices of Lagrange multipliers
to diagonal form would make the orbitals delocalized.
Hence, 
in problems where working with localized
orbitals is of advantage or relevance,
for instance in lattices,
it is the form (\ref{MCTDH_conver_stationary_orb})
with the in-general 
non-diagonal
Lagrange multipliers 
which is to 
be preferred.

\subsection{Formulation for general interactions}\label{non_contact}

The last stage of the theory is to 
return to the case of generic non-contact 
interactions in the Hamiltonian
(\ref{ham_am_1}-\ref{ham_am_3})
and have the respective 
theory derived.
The derivation of the equations of motion
follows essentially the same 
steps taken 
in the previous subsection \ref{contact}
and there is obviously no need to repeat it.
The only extra care needed
is when minimizing the 
conversion term 
in the 
functional action with respect to 
the molecular orbitals,
where exchange of variables is used.
This point and 
related derivations are 
discussed in appendix \ref{appen_C}.
Below,
we report the final results 
of the time-dependent 
as well as the self-consistent 
time-independent theories.

The final form of the time-dependent 
equations of motion of the orbitals reads,
$j=1,\ldots,M$, $j'=1,\ldots,M'$:
\beqn\label{MCTDH_W_conver_orb_eqs}
& & \!\!\!\!\!\!\!\! i\left|\dot\phi_j\right> =
   \hat {\mathbf P}^{(a)} \Bigg[\hat h^{(a)} \left|\phi_j\right> + \nonumber \\
 & & + \sum^M_{k=1} \left\{\brho^{(a)}(t)\right\}^{-1}_{jk} \times  \sum^M_{q=1}
\Bigg( \left\{\rho_2W\right\}^{(a)}_{kq} \left|\phi_q\right> 
 + \sqrt{2} \sum_{k'=1}^{M'} 
\rho^{(m\rightharpoondown 2a)}_{qkk'} \hat W^{(m\rightharpoondown 2a)}_{qk'}\Bigg) \Bigg], \nonumber \\
& & \!\!\!\!\!\!\!\! i\left|\dot\psi_{j'}\right> =
   \hat {\mathbf P}^{(m)} \Bigg[\hat h^{(m)} \left|\psi_{j'}\right> + \nonumber \\
 & & \sum^{M'}_{k'=1} \left\{\brho^{(m)}(t)\right\}^{-1}_{j'k'} \times
\Bigg( \sum^{M'}_{q'=1} \left\{\rho_2W\right\}^{(m)}_{k'q'}
\left|\psi_{q'}\right> 
 + \frac{1}{\sqrt{2}} \sum_{k,q=1}^{M} 
\rho^{(2a\rightharpoonup m)}_{kk'q} \hat W^{(2a\rightharpoonup m)}_{kq}\Bigg) \Bigg], \
\eeqn
where terms 
with products of 
reduced two-body density matrices 
times one-body potentials (see below) 
are collected together
and denoted 
for brevity as
\beqn\label{rho_W}
 & & \left\{\rho_2W\right\}^{(a)}_{kq} \equiv 
 \sum^M_{s,l=1} \rho^{(a)}_{kslq} \hat W^{(a)}_{sl} + 
\sum_{k',q'=1}^{M'} \rho^{(am)}_{kk'qq'} \hat W^{(am)}_{k'q'}, \nonumber \\
 & & \left\{\rho_2W\right\}^{(m)}_{k'q'} \equiv 
 \sum^{M'}_{s',l'=1} \rho^{(m)}_{k's'l'q'} \hat W^{(m)}_{s'l'} +
 \sum_{k,q=1}^{M} \rho^{(am)}_{kk'qq'} \hat W^{(ma)}_{kq}. \
\eeqn
Comparing Eq.~(\ref{MCTDH_conver_orb_eqs}) of the previous
subsection \ref{contact} to Eq.~(\ref{MCTDH_W_conver_orb_eqs}),
we see that in the latter more general, 
time-dependent local potentials appear
which are given explicitly by:
\beqn\label{one_body_pot}
 & & \hat W^{(a)}_{sl}(\r,t) =
\int \phi_s^\ast(\r',t) \hat W^{(a)}(\r,\r') \phi_l(\r',t) d\r', \nonumber \\
 & & \hat W^{(m)}_{s'l'}(\r,t) = 
\int \psi_{s'}^\ast(\r',t) \hat W^{(m)}(\r,\r') \psi_{l'}(\r',t) d\r', \nonumber \\
 & & \hat W^{(am)}_{k'q'}(\r,t) = 
\int \psi_{k'}^\ast(\r',t) \hat W^{(am)}(\r,\r') \psi_{q'}(\r',t) d\r', \nonumber \\
 & & \hat W^{(ma)}_{kq}(\r,t) = 
\int \phi_{k}^\ast(\r',t) \hat W^{(am)}(\r,\r') \phi_{q}(\r',t) d\r', \nonumber \\
 & & \hat W^{(2a\rightharpoonup m)}_{kq}(\r,t) =
\int d\r' \hat W^{(2a\rightharpoonup m)}\!\left(\r+\frac{\r'}{2},\r-\frac{\r'}{2}\right)
 \phi_k\!\left(\r+\frac{\r'}{2},t\right) \phi_q\!\left(\r-\frac{\r'}{2},t\right), \nonumber \\ 
 & & \hat W^{(m\rightharpoondown 2a)}_{qk'}(\r,t) = 
\int d\r' \phi_q^\ast(\r',t) \hat W^{(m\rightharpoondown 2a)}(\r,\r') 
\psi_{k'}\!\left(\frac{\r+\r'}{2},t\right). \
\eeqn
These potentials derive from the interaction terms 
and conversion term
in the Hamiltonian 
(\ref{ham_am_1}-\ref{ham_am_3}) and,
in the specific case of contact 
particle-particle
interactions,
boil down to products-of-orbitals,
see for comparison Eq.~(\ref{MCTDH_conver_orb_eqs}).
The form of the equations of motion
for the corresponding expansion coefficients,
\beqn\label{MCTDH_W_conver_coeff_eqs}
& & \qquad {\mathbf H}^{(2a \leftrightharpoons m)}(t)\C(t) = i\frac{\partial \C(t)}{\partial t}, \nonumber \\
& &  H^{(2a \leftrightharpoons m)}_{\vec{n}^p\vec{m}^p,\vec{n}'^{p'}\vec{m}'^{p'}}(t) =
\left<\vec{n}^p,\vec{m}^p;t\left|\hat H^{(2a \leftrightharpoons m)} \right|\vec{n}'^{p'},\vec{m}'^{p'};t\right>, \
\eeqn
does 
not change for general interactions.
Of course, the matrix 
elements $H^{(2a \leftrightharpoons m)}_{\vec{n}^p\vec{m}^p,\vec{n}'^{p'}\vec{m}'^{p'}}(t)$
do depend on the specific 
form of the 
particle-particle
interactions.

Finally, the self-consistent, time-independent 
general
theory is obtained from the time-dependent one
by taking $t \to -it$. 
The stationary self-consistent equations for the orbitals read:
\beqn\label{MCTDH_conver_stationary_orb_gen}
& & \!\!\!\!\!\!\! 
\sum_{q=1}^M \Bigg\{ \Bigg[ \rho^{(a)}_{kq}\hat h^{(a)} +
 \left\{\rho_2W\right\}^{(a)}_{kq} \Bigg] \left|\phi_q\right> 
+ \sqrt{2} \sum_{k'=1}^{M'} \rho^{(m\rightharpoondown 2a)}_{qkk'} \hat W^{(m\rightharpoondown 2a)}_{qk'} \Bigg\}
= \sum_{j=1}^M \mu^{(a)}_{kj} \left|\phi_j\right> \to \mu^{(a)}_{k} \left|\phi_k\right>, \nonumber \\
& & \qquad k=1,\ldots,M, \nonumber \\
& & \!\!\!\!\!\!\! 
\sum_{q'=1}^{M'} \Bigg[ \rho^{(m)}_{k'q'}\hat h^{(m)} +
 \left\{\rho_2W\right\}^{(m)}_{k'q'}
\Bigg] \left|\psi_{q'}\right> + \frac{1}{\sqrt{2}} \sum_{k,q=1}^{M} 
\rho^{(2a\rightharpoonup m)}_{k'kq} \hat W^{(2a\rightharpoonup m)}_{kq} 
= \sum_{j'=1}^{M'} \mu^{(m)}_{k'j'} \left|\psi_{j'}\right> \to \mu^{(m)}_{k'} 
\left|\psi_{k'}\right>, \nonumber \\
& & \qquad k'=1,\ldots,M', \
\eeqn
where the arrows indicate 
the Lagrange multipliers in their diagonal form,
as done in Eq.~(\ref{diagonal_form}).
Finally, 
the self-consistent--eigenvalue form 
of the equation for the expansion coefficients, 
\beqn\label{MCTDH_conver_stationary_coeff_gen}
 & & \qquad {\mathbf H}^{(2a\leftrightharpoons m)} \C = \varepsilon \C \nonumber \\
 & &  H^{(2a \leftrightharpoons m)}_{\vec{n}^p\vec{m}^p,\vec{n}'^{p'}\vec{m}'^{p'}} =
\left<\vec{n}^p,\vec{m}^p\left|\hat H^{(2a \leftrightharpoons m)} \right|\vec{n}'^{p'},\vec{m}'^{p'}\right>, \
\eeqn
remains unchanged 
for general interactions.

Eqs.~(\ref{MCTDH_W_conver_orb_eqs}) and 
(\ref{MCTDH_W_conver_coeff_eqs})
constitute a multiconfigurational {\it time-dependent} 
theory for systems of 
bosonic
atoms and molecules with conversion 
(particle conversion in the generic case).
Furthermore,
Eqs.~(\ref{MCTDH_conver_stationary_orb_gen}) and 
(\ref{MCTDH_conver_stationary_coeff_gen})
constitute a multiconfigurational self-consistent 
{\it time-independent} theory 
for systems of 
bosonic
atoms and molecules with conversion 
(particle conversion in the generic case).
Both theories extend the 
scope of the successful multiconfigurational
time-dependent Hartree method and its 
versions specified for systems of identical particles and mixtures
to new physical systems and problems.

\section{Summary and concluding remarks}\label{dis_sum}

In this work we have derived a many-body propagation theory 
for systems with particle conversion.
The theory is intended for systems with 
a finite number of interacting particles,
typically in a trap potential.
The theory has been exemplified 
and working 
equations have been explicitly derived for systems of 
interacting 
structureless bosonic atoms and bosonic molecules undergoing 
the conversion `reaction' $2a \leftrightharpoons m$.
In doing so,
we have also extended 
the scope of the successful multiconfigurational
time-dependent Hartree (MCTDH) method 
and its 
versions specified for systems of 
identical particles and mixtures
to new physical systems and problems.
We note that the MCTDH method is considered at present 
the most efficient
wave-packet propagation approach for in-general 
distinguishable coupled degrees-of-freedom,
with no particle conversion, 
of course.
The general multiconfigurational theory 
with particle conversion shall be referred to as
MCTDH-{\it conversion},
whereas the explicit scenario derived 
throughout this work by 
MCTDH-[$2a{\leftrightharpoons}m$].

To treat systems with particle conversion, 
one has to work in second quantization formalism,
where the Hamiltonian with particle-conversion terms 
can be represented.
The next step is to define the configurations.
In presence of particle conversion configurations with 
different numbers of
atoms and different numbers of molecules 
are coupled.
For instance, 
consider
the particular case of the conversion `reaction' 
$2a \leftrightharpoons m$.
In this case, 
the subspace of coupled configurations can be easily obtained
by
starting 
from
the configurations made of $N$ atoms only,
and operating 
repeatedly 
with the conversion operators in the Hamiltonian 
until configurations made of the maximal number $\left[\frac{N}{2}\right]$ 
of molecules are 
reached.

In the multiconfigurational theory
for the 'reaction' $2a \leftrightharpoons m$,
there are $M$ {\it time-dependent} 
orbitals $\{\phi_k(\r,t)\}$ available for the atoms and
$M'$ {\it time-dependent} orbitals $\{\psi_{k'}(\r,t)\}$ for the molecules.
The multiconfigurational ansatz for the many-particle
wavefunction $\Psi(t)$ is taken as linear combination 
with {\it time-dependent} coefficients $\{C_{\vec{n}^p\vec{m}^p}(t)\}$
of all possible configurations $\left\{\left|\vec{n}^p,\vec{m}^p;t\right>\right\}$ 
assembled from 
$p$ molecules and $N-2p$ atoms -- distributed over the $M$ and $M'$ respective orbitals -- 
and coupled by the conversion term in the Hamiltonian.
  
The evolution of $\Psi(t)$
is then determined by the Dirac-Frenkel 
time-dependent variational principle.
Utilizing the Lagrangian formulation 
of the Dirac-Frenkel variational principle, 
one arrives at two sets of coupled equations of motion:
The first set is for the orbitals $\{\phi_k(\r,t)\}$ and $\{\psi_{k'}(\r,t)\}$,
and the second for the expansion 
coefficients $\{C_{\vec{n}^p\vec{m}^p}(t)\}$.
The first set 
is comprised 
of first-order differential equations in time and non-linear
integrodifferential 
(for non-contact interactions)
in position space.
The second set consists of first-order differential equations 
with coefficients forming 
a time-dependent Hermitian matrix.
Thus,
equations of motion (\ref{MCTDH_conver_orb_eqs_P},\ref{MCTDH_conver_coeff_eqs_P}), 
or Eqs.~(\ref{MCTDH_conver_orb_eqs},\ref{MCTDH_conver_coeff_eqs}),
constitute the time-dependent multiconfigurational theory for 
bosonic atoms and molecules with conversion -- MCTDH-[$2a{\leftrightharpoons}m$].

The structure of the equations of motion for systems
with particle conversion reminds of 
the structure of the 
equations of motion in multiconfigurational time-dependent theories for
systems of identical particles and mixtures \cite{MCTDHB1,MCTDHB2,Unified_paper,MIX}:
(i) There are projection operators on the right-hand-sides of the equations of motion 
for the orbitals,
ensuring that the respective orbitals remain normalized and orthogonal to one another for all times; 
(ii) The equations for the expansion coefficients are first-order
differential equations with time-dependent coefficients; and
(iii) The equations of motion for the orbitals are 
formulated in terms of reduced density matrices.
This resemblance would allow one to transfer the 
effective numerical techniques
that have been developed 
in the past almost-twenty years
for multiconfigurational 
time-dependent many-body systems
{\it without} particle conversion 
\cite{CPL,JCP,PR,MCTDHF1,MCTDHF2,MCTDHF3,MCTDHB1,MCTDHB2,Unified_paper,MIX} 
to the present theory for 
systems {\it with} 
particle conversion.

Particular attention has been paid to
the reduced density matrices appearing in the theory.
As the multiconfigurational expansion 
involves configurations 
with different numbers of atoms and molecules,
two types of reduced density matrices are defined.
There are {\it particle-conserving}
reduced density matrices,
$\rho^{(a)}(\r_1|\r_2;t)$, $\rho^{(m)}(\r_1|\r_2;t)$, $\rho^{(a)}(\r_1,\r_2|\r_3,\r_4;t)$,
$\rho^{(m)}(\r_1,\r_2|\r_3,\r_4;t)$ and $\rho^{(am)}(\r_1,\r_2|\r_3,\r_4;t)$,
which directly do not couple configurations with different numbers
of atoms and molecules.
Despite this property,
the particle-conserving 
reduced density matrices, 
such as $\rho^{(a)}(\r_1|\r_2;t)$ and $\rho^{(a)}(\r_1,\r_2|\r_3,\r_4;t)$,
{\it are not} the standard density matrices 
introduced by L\"owdin \cite{Lowdin} 
for many-particle systems {\it without} conversion.
The second type of reduced density matrices that appear in 
the theory are 
{\it particle non-conserving} 
reduced density matrices,
$\rho^{(a\rightharpoonup m)}(\r_1,\r_2|\r_3;t)$ and $\rho^{(m\rightharpoondown a)}(\r_3|\r_2,\r_1;t)$,
and originate from the conversion term in the Hamiltonian.
They obviously have no analogs in systems without conversion.
Here,
it is of interest by itself to study
properties of particle-conserving 
reduced density matrices and certainly
of particle non-conserving reduced density matrices 
in systems {\it with} particle conversion.

The time-dependent multiconfigurational theory 
MCTDH-[$2a{\leftrightharpoons}m$] readily admits
the corresponding {\it stationary} theory.
By resorting to imaginary time propagation, 
the equations of motion of the time-dependent
theory boil down to 
the fully-self-consistent time-independent 
multiconfigurational theory,
Eqs.~(\ref{MCTDH_conver_stationary_orb_gen},\ref{MCTDH_conver_stationary_coeff_gen}), 
for stationary states of the system $2a \leftrightharpoons m$,
in presence of all particle-particle interactions, 
of course.
With this result,
available self-consistent multiconfigurational
theories for systems 
{\it without} 
particle conversion,
noticeably for fermions \cite{Szabo_book,elect}, 
distinguishable degrees-of-freedom \cite{Dieter_review}, 
bosons \cite{MCHB}, 
and mixtures \cite{MIX},
are taken a step further,
to systems 
{\it with} 
particle conversion. 

A specific case of interest for systems 
of bosonic atoms and molecules with conversion
is the case of $M=1$ atomic and $M'=1$ molecular orbitals,
which is presented in section \ref{two_mode_AM}
before the general MCTDH-[$2a{\leftrightharpoons}m$]
theory
is developed.
For $M=1$ and $M'=1$,
the corresponding multiconfigurational theory
is the fully-variational theory 
that results when 
the shape of the atomic $\phi_a(\r,t)$
and molecular $\psi_m(\r,t)$
orbitals
and of 
each and every expansion
coefficient $C_p(t)$
are optimized according to the variational principle.
Being fully-variational with respect to the shape
of the orbitals $\phi_a(\r,t)$, $\psi_m(\r,t)$
and with respect to the expansion 
coefficients $\{C_p(t)\}$,
the theory generalizes the literature 
Gross-Pitaevskii equation \cite{Timmermans_review} and two-mode 
approximation \cite{2M1,Vardi} for bosonic atoms and molecules with conversion.
We term 
this 
specific case of the general theory {\it conversion mean field},
as there is only one orbital available for the
bosonic atoms and one for the molecules --
the minimal number possible 
for bosonic species.

At the other end,
in the limit where the number $M$ of atomic orbitals 
and $M'$ of molecular orbitals goes to infinity,
the MCTDH-[$2a{\leftrightharpoons}m$] theory becomes an
exact representation of the time-dependent
many-particle Schr\"odinger equation 
with the particle-conversion 
Hamiltonian $\hat H^{(2a\rightleftharpoons m)}$.
In practice,
one obviously has to limit $M$ and $M'$.
Here,
the employment of {\it time-dependent} orbitals,
which has been very successful for
the MCTDH approach and its 
versions 
specified
for identical particles and mixtures,
is of great help and advantage.
Of course,
even with time-dependent orbitals
the size of the Hilbert space grows rapidly
with the 
size of the 
system
and the number of orbitals $M$ and $M'$
employed.   
Consequently,
with increasing system size 
and 
as the number of orbitals 
which
one has to employ becomes larger,
e.g., for stronger interactions,
it is instructive to 
devise
truncation schemes beyond the
usage of {\it time-dependent} multiconfigurational 
expansions over 
{\it complete} 
Hilbert subspaces.
We mention two such truncation strategies:
(i) to truncate time-dependent multiconfigurational expansions
to include parts of Hilbert subspaces,
i.e., to include not all available configurations
for a given 
system size
and number of orbitals $M$,$M'$; 
and
(ii) to concentrate on the reduced density matrices,
write equations of motion for them directly,
and thereafter truncate the resulting hierarchy 
of equations of motion for higher-order reduced density matrices
at some given order.
The development of these truncation schemes for 
{\it time-dependent} multiconfigurational expansions
in systems with particle conversion 
extends beyond the scope of the present work.

Finally,
the explicit equations of motion presented in this work are
for the specific `reaction' $2a \leftrightharpoons m$
where the atoms and molecules are structureless bosons --
the MCTDH-[$2a{\leftrightharpoons}m$] theory. 
Several other systems come to mind:
(i) Other `reactions' with bosonic atoms of the same kind, 
e.g., $3a \leftrightharpoons m$;
(ii) `Reactions' with bosonic atoms of a different kind, 
e.g., $a+a' \leftrightharpoons m$.
In this case and for general 
particle-particle interactions,
the center-of-mass coordinate is, of course, 
$\R=\frac{m_a\r+m_{a'}\r'}{m_a+m_{a'}}$,
where $m_a$ and $m_{a'}$ are the masses of the respective species;
(iii) `Reactions' including fermionic atoms, 
e.g., $a_f+a_b \leftrightharpoons m_f$ and $a_f+a_f \leftrightharpoons m_b$
where the subscript $b$, $f$ stands for bosonic, fermionic species.
In the latter case, 
a unified form 
of the respective equations of motion
and those of the present work is anticipated; 
and 
(iv) A whole zoo of `reactions' for particles with spin, 
internal-structure. 
The extension of MCTDH-[$2a{\leftrightharpoons}m$]
for the above concrete examples 
as well as for other systems 
with particle conversion 
can be done by following the 
theory and derivation steps
of the present work.

\begin{acknowledgments}
Financial support by the Deutsche Forschungsgemeinschaft 
is gratefully acknowledged.
\end{acknowledgments}

\appendix

\section{Further details of the derivation of the equations of motion}\label{Lagrange}

\subsection{The specific case of the 
conversion mean field 
(fully-variational two-mode approximation)}\label{Lagrange_two_mode}

When the variations of the functional action (\ref{action_functional_AM})
with respect to the orbitals and expansion coefficients are put to zero,
the following equations of motion are obtained:
\beqn\label{general_EOM_orbitals}
& & \left[\left<\hat N_a\right> \left(\hat h^{(a)} - i\frac{\partial}{\partial t}\right) +
\lambda_a \left<\hat N_a(\hat N_a-1)\right> |\phi_a|^2 + 
 \lambda_{am} \left<\hat N_a \hat N_m\right> |\psi_m|^2 \right] \left|\phi_a\right> + \nonumber \\
 & & \qquad + \sqrt{2}\lambda_{con} \left<\hat b_a^\dag \hat b_a^\dag \hat c_m\right>
 \phi_a^\ast \left|\psi_m\right> = 
 \mu_a(t) \left|\phi_a\right>, 
\nonumber \\
& & \left[\left<\hat N_m\right> \left(\hat h^{(m)} - i\frac{\partial}{\partial t}\right) +
\lambda_m \left<\hat N_m(\hat N_m-1)\right> |\psi_m|^2 +
 \lambda_{am} \left<\hat N_a \hat N_m\right> |\phi_a|^2 \right] \left|\psi_m\right> + \nonumber \\
 & & \qquad + \frac{\lambda_{con}}{\sqrt{2}} 
 \left<\hat c_m^\dag \hat b_a \hat b_a\right>  
\phi_a \left|\phi_a\right> = 
 \mu_m(t) \left|\psi_m\right> \
\eeqn
and
\beq\label{general_EOM_coeffieints}
 \qquad \left[\bcalH^{(a \leftrightharpoons m)}(t) - \varepsilon(t) \cdot \1 \right] 
\C(t) = i \frac{\partial\C(t)}{\partial t}. \ 
\eeq
The three time-dependent Lagrange multipliers
$\mu_a(t)$, $\mu_m(t)$ and $\varepsilon(t)$
appear therein.
How to eliminate them?

It is straightforward to eliminate $\varepsilon(t)$.
This is done by transforming the expansion coefficients as follows:
\beq\label{C_coeff_global_phase}
  \overline{\C}(t) = e^{-i\int \varepsilon(t') dt'} \C(t).
\eeq
Substituting Eq.~(\ref{C_coeff_global_phase}) into (\ref{general_EOM_coeffieints}) 
and removing at the end the 'bar' from all quantities we obtain: 
\beq\label{general_simplest_EOM_coeffieints}
\bcalH^{(a \leftrightharpoons m)}(t) \C(t) = i \frac{\partial\C(t)}{\partial t}.
\eeq
From Eqs.~(\ref{general_EOM_coeffieints}-\ref{general_simplest_EOM_coeffieints}) 
we see that 
the role of the Lagrange multiplier $\varepsilon(t)$ 
is that of a (redundant) global time-dependent phase 
of the many-particle wavefunction 
$\overline{\Psi}(t) = e^{-i\int \varepsilon(t') dt'}\Psi(t)$.
We note that Eq.~(\ref{general_EOM_orbitals}) is not affected by the transformation 
(\ref{C_coeff_global_phase}) because  
reduced density matrices 
are ``insensitive'' 
to a global phase of the wavefunction,
namely
$\left<\overline{\Psi}(t)\left| \, \ldots \, \right|\overline{\Psi}(t)\right> =
\left<\Psi(t)\left| \, \ldots \, \right|\Psi(t)\right>$.

The next step is 
to eliminate the remaining Lagrange multipliers $\mu_a(t)$ and $\mu_m(t)$.
Making use of the orbitals being normalized and 
taking the respective scalar products of Eq.~(\ref{general_EOM_orbitals})
with $\left<\phi_a\right|$ and $\left<\psi_m\right|$,
we obtain:
\beqn\label{Lagrange_2orbitals}
\mu_a(t) &=& \left<\hat N_a\right> \left<\phi_a\left|\hat h^{(a)} - i\frac{\partial}{\partial t}\right|\phi_a\right> +
 \frac{\lambda_a}{2} \left<\hat N_a(\hat N_a-1)\right> \left<\phi_a^2\left|\right.\phi_a^2\right> + \nonumber \\
 &+& \lambda_{am} \left<\hat N_a \hat N_m\right> \left<\phi_a\psi_m\left|\right.\phi_a\psi_m\right> + 
 \sqrt{2}\lambda_{con} \left<\hat b_a^\dag \hat b_a^\dag \hat c_m\right> \left<\phi_a^2\left|\right.\psi_m\right>, \nonumber \\
\mu_m(t) &=& \left<\hat N_m\right> \left<\phi_m\left|\hat h^{(m)} - i\frac{\partial}{\partial t}\right|\phi_m\right> +
 \frac{\lambda_m}{2} \left<\hat N_m(\hat N_m-1)\right> \left<\phi_m^2\left|\right.\phi_m^2\right> + \nonumber \\
 &+& \lambda_{am} \left<\hat N_a \hat N_m\right> \left<\phi_a\psi_m\left|\right.\phi_a\psi_m\right> + 
 \frac{\lambda_{con}}{\sqrt{2}} \left<\hat c_m^\dag \hat b_a \hat b_a\right> \left<\psi_m\left|\right.\phi_a^2\right>. \
\eeqn
Substituting 
Eq.~(\ref{Lagrange_2orbitals}) 
into Eq.~(\ref{general_EOM_orbitals}),
employing the identities:
\beqn\label{mu_identities}
& & \bigg[\left<\hat N_a\right> \left(\hat h^{(a)} - i\frac{\partial}{\partial t}\right) +
\lambda_a \left<\hat N_a(\hat N_a-1)\right> |\phi_a|^2 + \nonumber \\
 & & \qquad + \lambda_{am} \left<\hat N_a \hat N_m\right> |\psi_m|^2 \bigg] \left|\phi_a\right> + 
  \sqrt{2}\lambda_{con} \left<\hat b_a^\dag \hat b_a^\dag \hat c_m\right>
 \phi_a^\ast \left|\psi_m\right> - \mu_a(t) \left|\phi_a\right> = \nonumber \\ 
 & & = \left(1 - \left|\phi_a\left>\right<\phi_a\right|\right) 
\Bigg\{\bigg[\left<\hat N_a\right> \left(\hat h^{(a)} - i\frac{\partial}{\partial t}\right) +
\lambda_a \left<\hat N_a(\hat N_a-1)\right> |\phi_a|^2 + \nonumber \\
 & & \qquad + \lambda_{am} \left<\hat N_a \hat N_m\right> |\psi_m|^2 \bigg] \left|\phi_a\right> + 
 \sqrt{2}\lambda_{con} \left<\hat b_a^\dag \hat b_a^\dag \hat c_m\right>
 \phi_a^\ast \left|\psi_m\right>\Bigg\}, \nonumber \\
& & \bigg[\left<\hat N_m\right> \left(\hat h^{(m)} - i\frac{\partial}{\partial t}\right) +
\lambda_m \left<\hat N_m(\hat N_m-1)\right> |\psi_m|^2 + \nonumber \\
  & & \qquad + \lambda_{am} \left<\hat N_a \hat N_m\right> |\phi_a|^2 \bigg] \left|\psi_m\right> + 
 \frac{\lambda_{con}}{\sqrt{2}} 
 \left<\hat c_m^\dag \hat b_a \hat b_a\right>  
\phi_a \left|\phi_a\right> - \mu_m(t) \left|\psi_m\right> = \nonumber \\
 & & = \left(1-\left|\psi_m\left>\right<\psi_m\right|\right)
\Bigg\{\bigg[\left<\hat N_m\right> \left(\hat h^{(m)} - i\frac{\partial}{\partial t}\right) +
\lambda_m \left<\hat N_m(\hat N_m-1)\right> |\psi_m|^2 + \nonumber \\
  & & \qquad + \lambda_{am} \left<\hat N_a \hat N_m\right> |\phi_a|^2 \bigg] \left|\psi_m\right> +
 \frac{\lambda_{con}}{\sqrt{2}} 
 \left<\hat c_m^\dag \hat b_a \hat b_a\right>  
\phi_a \left|\phi_a\right>\Bigg\},   
\eeqn
and dividing the result, respectively, 
by $\left<\hat N_a\right>$ and 
$\left<\hat N_m\right>$,
the equations of motion 
(\ref{simplest_EOM_orbitals}) 
are obtained.

To eliminate the projection operators 
in front of the time derivatives 
in Eq.~(\ref{mu_identities}),
i.e., $\hat{\mathbf P}^{(a)} = 1 - \left|\phi_a\left>\right<\phi_a\right|$ 
and $\hat{\mathbf P}^{(m)} = 1 - \left|\psi_m\left>\right<\psi_m\right|$
on the left-hand sides of Eq.~(\ref{simplest_EOM_orbitals}),
we exploit the invariance property of the wavefunction $\Psi(t)$.
Consider the following phase transformations of the orbitals and coefficients:
\beqn\label{trans_appen_orb_coeff}
 & & \overline{\phi}_a(\r,t) = e^{+i\beta_a(t)} \phi_a(\r,t), \qquad 
\overline{\psi}_m(\r,t) =  e^{+i\gamma_m(t)} \psi_m(\r,t), \nonumber \\
 & & \overline{C}_p(t) = e^{-i\left[(N-2p) \beta_a(t) + p \gamma_m(t)\right]} C_p(t), 
      \qquad p=0,\ldots,\left[N/2\right]. \ 
\eeqn
Combing these phase transformations,
the wavefunction does not change:
$\left|\Psi(t)\right> = \sum_{p=0}^{\left[N/2\right]} C_p(t) \left|N-2p,p;t\right> =
 \sum_{p=0}^{\left[N/2\right]} \overline{C}_p(t) \overline{\left|N-2p,p;t\right>}$. 
We should also recall that transforming the orbitals 
goes along with transforming the corresponding annihilation, creation 
operators
(another way to look at this is that the field 
operators $\hat{\mathbf \Psi}_a(\r)$ and $\hat{\mathbf \Psi}_m(\r)$
are time-independent, 
basis-set-independent 
quantities and, 
consequently, 
transforming the orbitals
requires the reverse transformation 
of the annihilation operators).
Thus,
Eq.~(\ref{trans_appen_orb_coeff}) 
implies also the phase transformations
\beq\label{trans_appen_anni}
 \hat{\overline{b}}_a(t) = e^{-i\beta_a(t)} \hat b_a(t), \qquad 
 \hat{\overline{c}}_m(t) = e^{-i\gamma_m(t)} \hat c_m(t)
\eeq
for the atomic and molecular annihilation operators.

Now, 
plugging Eqs.~(\ref{trans_appen_orb_coeff}-\ref{trans_appen_anni})
into the equations of motion 
(\ref{simplest_EOM_orbitals}) for the orbitals 
and (\ref{simplest_EOM_coeffieints}) for the expansion coefficients
[see also Eqs.~(\ref{general_simplest_EOM_coeffieints},\ref{mu_identities})],
and choosing the phases
\beq\label{phase_choice_1}
 \beta_a(t) =  \int i \left<\phi_a(t')\left|\right.\dot\phi_a(t')\right> dt', \qquad 
 \gamma_m(t) = \int i \left<\psi_m(t')\left|\right.\dot\psi_m(t')\right> dt',
\eeq
equations of motion 
(\ref{simplest_EOM_orbitals_final}) 
and (\ref{simplest_EOM_coeffieints_final})
are found.
We note that the phases 
$\beta_a(t)$ and $\gamma_m(t)$ are real quantities
since $\left<\phi_a(t)\left|\right.\phi_a(t)\right>=1$
and $\left<\psi_m(t)\left|\right.\psi_m(t)\right>=1$,
respectively,
for all times.

\subsection{The general multiconfigurational theory}\label{appen_C}

When the expressions for the field operators
$\hat{\mathbf \Psi}_a(\r)$ and $\hat{\mathbf \Psi}_m(\r)$
in terms of the time-dependent orbitals, Eq.~(\ref{annihilation_def}),
are substituted into the generic many-body Hamiltonian (\ref{ham_am_1}-\ref{ham_am_3}),
one obtains:
\beqn\label{ham_general_inter}
& & 
 \hat H^{(2a\rightleftharpoons m)} = 
\sum_{k,q} h^{(a)}_{kq} \hat b_k^\dag \hat b_q + 
\sum_{k,s,l,q} W^{(a)}_{ksql} \hat b_k^\dag \hat b_s^\dag \hat b_l \hat b_q + 
\sum_{k',q'} h^{(m)}_{k'q'} \hat c_{k'}^\dag \hat c_{q'} + 
\sum_{k',s',l',q'} W^{(m)}_{k's'q'l'} \hat c_{k'}^\dag \hat c_{s'}^\dag \hat c_{l'} \hat c_{q'} + \nonumber \\
& & \qquad + \sum_{k,k',q,q'} W^{(am)}_{kk'qq'} \hat b_k^\dag \hat b_q \hat c_{k'}^\dag \hat c_{q'} + 
 \frac{1}{\sqrt{2}} 
\sum_{k',k,q} \left[W^{(2a\rightharpoonup m)}_{k'kq} \hat c_{k'}^\dag \hat b_k \hat b_q +
  W^{(m\rightharpoondown 2a)}_{qkk'} \hat b_q^\dag \hat b_k^\dag \hat c_{k'}
\right]. \
\eeqn
The one-body, 
two-body 
and conversion 
matrix elements appearing 
in Eq.~(\ref{ham_general_inter}) are given by:
\beqn\label{one_two_matrix_elements}
\!\!\!\!\!\!\!\!\!\!\! h^{(a)}_{kq} &=& \int \phi_k^\ast(\r,t) \hat h^{(a)}(\r) \phi_q(\r,t) d\r, \nonumber \\
\!\!\!\!\!\!\!\!\!\!\! W^{(a)}_{ksql} &=& \int \!\! \int \phi_k^\ast(\r,t) \phi_s^\ast(\r',t) \hat W^{(a)}(\r,\r')
 \phi_q(\r,t) \phi_l(\r',t) d\r d\r', \nonumber \\
\!\!\!\!\!\!\!\!\!\!\! h^{(m)}_{k'q'} &=&
\int \psi_{k'}^\ast(\r,t) \hat h^{(m)}(\r) \psi_{q'}(\r,t) d\r, \nonumber \\
\!\!\!\!\!\!\!\!\!\!\! W^{(m)}_{k's'q'l'}
&=& \int \!\! \int \psi_{k'}^\ast(\r,t) \psi_{s'}^\ast(\r',t) \hat W^{(m)}(\r,\r')
 \psi_{q'}(\r,t) \psi_{l'}(\r',t) d\r d\r', \nonumber \\
\!\!\!\!\!\!\!\!\!\!\! W^{(am)}_{kk'qq'}
&=& \int \!\! \int \phi_k^\ast(\r,t) \psi_{k'}^\ast(\r',t) \hat W^{(am)}(\r,\r')
 \phi_q(\r,t) \psi_{q'}(\r',t) d\r d\r', \nonumber \\
\!\!\!\!\!\!\!\!\!\!\! W^{(2a\rightharpoonup m)}_{k'kq} 
&=& \int \!\! \int \psi^\ast_{k'}\!\left(\frac{\r+\r'}{2},t\right) \hat W^{(2a\rightharpoonup m)}(\r,\r')
 \phi_k(\r,t) \phi_q(\r',t) d\r d\r' = \nonumber \\
\!\!\!\!\!\!\!\!\!\!\! &=& \int \!\! \int \psi^\ast_{k'}(\r,t) 
\hat W^{(2a\rightharpoonup m)}\!\left(\r+\frac{\r'}{2},\r-\frac{\r'}{2}\right)
 \phi_k\!\left(\r+\frac{\r'}{2},t\right) \phi_q\!\left(\r-\frac{\r'}{2},t\right) d\r d\r', \nonumber \\
\!\!\!\!\!\!\!\!\!\!\! W^{(m\rightharpoondown 2a)}_{qkk'} &=&
\int \!\! \int \phi^\ast_q(\r,t) \phi^\ast_k(\r',t) \hat W^{(m\rightharpoondown 2a)}(\r,\r')
\psi_{k'}\!\left(\frac{\r+\r'}{2},t\right) d\r d\r' = \left\{W^{(2a\rightharpoonup m)}_{k'kq}\right\}^\ast\!\!. \ 
\eeqn
The change of variables used
for $W^{(2a\rightharpoonup m)}_{k'kq}$
is needed in order to 
perform the variation of this term 
with respect to the molecular orbitals,
see below.

Now, 
the expectation value appearing in the functional 
action (\ref{action_functional_AM_full}) 
when expressed explicit 
with respect to the orbitals reads:
\beqn\label{exp_full_app}
& & \left<\Psi(t)\left| \hat H^{(2a\leftrightharpoons m)} - i\frac{\partial}{\partial t}\right|\Psi(t)\right> 
= \sum_{k,q=1}^M \rho^{(a)}_{kq} \left[ h^{(a)}_{kq} - {\left(i\frac{\partial}{\partial t}\right)}_{kq}^{\!\!(a)} \right] 
+ \frac{1}{2}\sum_{k,s,l,q=1}^M \rho^{(a)}_{kslq} W^{(a)}_{ksql} + \nonumber \\
& & + \sum_{k',q'=1}^{M'} \rho^{(m)}_{k'q'} \left[ h^{(m)}_{k'q'}
- {\left(i\frac{\partial}{\partial t}\right)}_{k'q'}^{\!\!(m)} \right]
 + \frac{1}{2}\sum_{k',s',l',q'=1}^{M'} \rho^{(m)}_{k's'l'q'} W^{(m)}_{k's'q'l'} + 
 \sum_{k,q=1}^M \sum_{k',q'=1}^{M'} \rho^{(am)}_{kk'qq'} W^{(am)}_{kk'qq'} + \nonumber \\
& & + \frac{1}{\sqrt{2}} \sum_{k,q=1}^{M} \sum_{k'=1}^{M'} 
\left[
\rho_{k'kq}^{(2a\rightharpoonup m)} W^{(2a\rightharpoonup m)}_{k'kq} +
\rho_{qkk'}^{(m\rightharpoondown 2a)} W^{(m\rightharpoondown 2a)}_{qkk'} \right] 
- i  \sum_{p=0}^{\left[N/2\right]} \sum_{\vec{n}^p,\vec{m}^p}
 C_{\vec{n}^p\vec{m}^p}^\ast \frac{\partial C_{\vec{n}^p\vec{m}^p}}{\partial t}, \ 
\eeqn
where
\beq
\left(i\frac{\partial}{\partial t}\right)_{kq}^{\!\!(a)} =
i\int \phi_k^\ast(\r,t) \frac{\partial\phi_q(\r,t)}{\partial t} d\r, \qquad
\left(i\frac{\partial}{\partial t}\right)_{k'q'}^{\!\!(m)} =
i\int \psi_{k'}^\ast(\r,t) \frac{\partial\psi_{q'}(\r,t)}{\partial t} d\r. 
\eeq
Equating 
the variation 
of the functional action (\ref{action_functional_AM_full})
with respect to the orbitals to zero, 
making use of Eq.~(\ref{exp_full_app}), 
the following equations are obtained:  
\beqn\label{general_variation_orbitals_direct}
& & \sum_{q=1}^M \Bigg\{ \Bigg[ \rho^{(a)}_{kq} 
\left(\hat h^{(a)} - i\frac{\partial}{\partial t}\right) +
  \sum^M_{s,l=1}\rho^{(a)}_{kslq} \hat W^{(a)}_{sl} + 
 \sum_{k',q'=1}^{M'} \rho^{(am)}_{kk'qq'} \hat W^{(am)}_{k'q'} \Bigg] \left|\phi_q\right> + \nonumber \\
& & \qquad + \sqrt{2} \sum_{k'=1}^{M'} \rho^{(m\rightharpoondown 2a)}_{qkk'} \hat W^{(m\rightharpoondown 2a)}_{qk'} \Bigg\}
= \sum_{j=1}^M \mu^{(a)}_{kj} \left|\phi_j\right>, \qquad k=1,\ldots,M, \nonumber \\
& & \sum_{q'=1}^{M'} \Bigg[ \rho^{(m)}_{k'q'} \left(\hat h^{(m)} - i\frac{\partial}{\partial t}\right) +
 \sum^{M'}_{s',l'=1}\rho^{(m)}_{k's'l'q'} \hat W^{(m)}_{s'l'} + 
 \sum_{k,q=1}^{M} \rho^{(am)}_{kk'qq'} \hat W^{(ma)}_{kq} \Bigg] \left|\psi_{q'}\right> + \nonumber \\
& & \qquad + \frac{1}{\sqrt{2}} \sum_{k,q=1}^{M} 
\rho^{(2a\rightharpoonup m)}_{k'kq} \hat W^{(2a\rightharpoonup m)}_{kq}
= \sum_{j'=1}^{M'} \mu^{(m)}_{k'j'} \left|\psi_{j'}\right>, 
\qquad k'=1,\ldots,M'. \
\eeqn
One delicate point in performing the variation of $W^{(2a\rightharpoonup m)}_{k'kq}$ 
[the first term in the last line of the expectation value Eq.~(\ref{exp_full_app})]
with respect to the molecular orbitals 
$\psi^\ast_{k'}(\r,t)$ 
is worth mentioning.
To perform this variation,
a change of the integration variables $\r$, $\r'$ 
to the center-of-mass $\R=\frac{\r+\r'}{2}$ and relative $\bar\r=\r-\r'$ coordinates is required.
Assigning thereafter back $\R \to \r$, $\bar\r \to \r'$, 
the matrix element $W^{(2a\rightharpoonup m)}_{k'kq}$ is re-written 
in a form, see Eq.~(\ref{one_two_matrix_elements}),
which is amenable to 
explicit variation with respect to 
$\psi^\ast_{k'}(\r,t)$.

The next step it to eliminate the Lagrange multipliers $\{\mu_{kj}^{(a)}(t)\}$ 
and $\{\mu_{k'j'}^{(m)}(t)\}$.
Making use of the orthonormality properties of the 
atomic and molecular
orbitals,
$\left<\phi_k(t)\left|\right.\phi_q(t)\right>=\delta_{kq}$ and 
$\left<\psi_{k'}(t)\left|\right.\psi_{q'}(t)\right>=\delta_{k'q'}$,
and taking the 
corresponding 
scalar products of Eq.~(\ref{general_variation_orbitals_direct})
with respect to the orbitals,
we obtain 
explicit expressions
for the 
Lagrange multipliers:
\beqn\label{MCTDH_conver_mu_gener}
 & & \mu_{kj}^{(a)}(t) =
 \sum^M_{q=1}\Bigg\{\rho_{kq}^{(a)} \left[h^{(a)}_{jq} - \left(i\frac{\partial}{\partial t}\right)_{jq}^{\!\!(a)} \right] +
  \sum^M_{s,l=1} \rho_{kslq}^{(a)} W^{(a)}_{jsql} + \nonumber \\
 & & \qquad + \sum_{k',q'=1}^{M'} \rho^{(am)}_{kk'qq'} W^{(am)}_{jk'qq'}
 + \sqrt{2}
\sum_{k'=1}^{M'} \rho^{(m\rightharpoondown 2a)}_{qkk'} W^{(m\rightharpoondown 2a)}_{qjk'} \Bigg\}, \nonumber \\
 & & \mu_{k'j'}^{(m)}(t) =
 \sum^{M'}_{q'=1}\Bigg\{\rho_{k'q'}^{(m)} \left[h^{(m)}_{j'q'} - 
\left(i\frac{\partial}{\partial t}\right)_{j'q'}^{\!\!(m)} \right] +
   \sum^{M'}_{s',l'=1} \rho_{k's'l'q'}^{(m)} W^{(m)}_{j's'q'l'} + \nonumber \\
 & & \qquad + \sum_{k,q=1}^{M} \rho^{(am)}_{kk'qq'} W^{(am)}_{kj'qq'} \Bigg\}
 + \frac{1}{\sqrt{2}} \sum_{k,q=1}^{M} 
 \rho^{(2a\rightharpoonup m)}_{k'kq} W^{(2a\rightharpoonup m)}_{j'kq}. \
\eeqn
Substituting Eq.~(\ref{MCTDH_conver_mu_gener})
into Eq.~(\ref{general_variation_orbitals_direct}),
making use of the identities:
\beqn\label{general_identities}
& & \!\!\!\!\!\!\!\!\!\! \sum_{q=1}^M \Bigg\{ \Bigg[ \rho^{(a)}_{kq} 
\left(\hat h^{(a)} - i\frac{\partial}{\partial t}\right) +
  \sum^M_{s,l=1}\rho^{(a)}_{kslq} \hat W^{(a)}_{sl} + 
 \sum_{k',q'=1}^{M'} \rho^{(am)}_{kk'qq'} \hat W^{(am)}_{k'q'} \Bigg] \left|\phi_q\right> + \nonumber \\
& &  \!\!\!\!\!\!\!\!\!\!
\qquad + \sqrt{2} \sum_{k'=1}^{M'} \rho^{(m\rightharpoondown 2a)}_{qkk'} \hat W^{(m\rightharpoondown 2a)}_{qk'} \Bigg\}
 - \sum_{u=1}^M \mu^{(a)}_{ku} \left|\phi_u\right> = \nonumber \\
& &  \!\!\!\!\!\!\!\!\!\!
\left(1-\sum_{u=1}^M \left|\phi_u\left>\right<\phi_u\right|\right) \sum_{q=1}^M \Bigg\{ \Bigg[ \rho^{(a)}_{kq} 
\left(\hat h^{(a)} - i\frac{\partial}{\partial t}\right) +
  \sum^M_{s,l=1}\rho^{(a)}_{kslq} \hat W^{(a)}_{sl} + \nonumber \\ 
& &  \!\!\!\!\!\!\!\!\!\! 
\qquad + \sum_{k',q'=1}^{M'} \rho^{(am)}_{kk'qq'} \hat W^{(am)}_{k'q'} \Bigg] \left|\phi_q\right> +
 \sqrt{2} \sum_{k'=1}^{M'} \rho^{(m\rightharpoondown 2a)}_{qkk'} 
\hat W^{(m\rightharpoondown 2a)}_{qk'} \Bigg\}, \qquad  k=1,\ldots,M, \nonumber \\
& &  \!\!\!\!\!\!\!\!\!\!
\sum_{q'=1}^{M'} \Bigg[ \rho^{(m)}_{k'q'} \left(\hat h^{(m)} - i\frac{\partial}{\partial t}\right) +
 \sum^{M'}_{s',l'=1}\rho^{(m)}_{k's'l'q'} \hat W^{(m)}_{s'l'} + 
 \sum_{k,q=1}^{M} \rho^{(am)}_{kk'qq'} \hat W^{(ma)}_{kq} \Bigg] \left|\psi_{q'}\right> + \nonumber \\
& & \!\!\!\!\!\!\!\!\!\! 
\qquad + \frac{1}{\sqrt{2}} \sum_{k,q=1}^{M} 
\rho^{(2a\rightharpoonup m)}_{k'kq} \hat W^{(2a\rightharpoonup m)}_{kq}
 - \sum_{u'=1}^{M'} \mu^{(m)}_{k'u'} \left|\psi_{u'}\right> =  \nonumber \\
& &  \!\!\!\!\!\!\!\!\!\!
 \left(1-\sum_{u'=1}^{M'} \left|\psi_{u'}\left>\right<\psi_{u'}\right|\right)
 \Bigg\{ \sum_{q'=1}^{M'} \Bigg[ \rho^{(m)}_{k'q'} \left(\hat h^{(m)} - i\frac{\partial}{\partial t}\right) +
 \sum^{M'}_{s',l'=1}\rho^{(m)}_{k's'l'q'} \hat W^{(m)}_{s'l'} + \nonumber \\
& &  \!\!\!\!\!\!\!\!\!\! 
\qquad + \sum_{k,q=1}^{M} \rho^{(am)}_{kk'qq'} \hat W^{(ma)}_{kq} \Bigg] \left|\psi_{q'}\right> + 
  \frac{1}{\sqrt{2}} \sum_{k,q=1}^{M} 
\rho^{(2a\rightharpoonup m)}_{k'kq} \hat W^{(2a\rightharpoonup m)}_{kq} \Bigg\}, \qquad k'=1,\ldots,M', \
\eeqn
and multiplying the result, respectively, by the
inverse of the reduced one-body density matrices 
and summing over
$\sum_{k=1}^M \left\{\brho^{(a)}(t)\right\}_{jk}^{-1}$ and
$\sum_{k'=1}^{M'} \left\{\brho^{(m)}(t)\right\}_{j'k'}^{-1}$,
we obtain equations of motion like (\ref{MCTDH_conver_orb_eqs_P})
with general interactions.

Finally,
to eliminate the projection operators 
$\hat{\mathbf P}^{(a)} = 1 - \sum_{u=1}^M \left|\phi_u\left>\right<\phi_u\right|$ 
and 
$\hat{\mathbf P}^{(m)} = 1 - \sum_{u'=1}^{M'} \left|\psi_{u'}\left>\right<\psi_{u'}\right|$
in front of the time derivatives 
[see Eqs.~(\ref{general_identities}) and (\ref{MCTDH_conver_orb_eqs_P})],
we employ the invariance properties of the 
wavefunction $\Psi(t)=\overline{\Psi}(t)$,
where 
$\left\{\phi_k(\r,t)\right\} \to \left\{\overline{\phi}_k(\r,t)\right\}$, 
$\left\{\psi_{k'}(\r,t)\right\} \to \left\{\overline{\psi}_{k'}(\r,t)\right\}$,
and
$\{C_{\vec{n}^p\vec{m}^p}(t)\} \to \{\overline{C}_{\vec{n}^p\vec{m}^p}(t)\}$.
For this,
consider the following time-dependent matrices:
\beq\label{matrix_U}
 \D^{(a)}(t), \ D^{(a)}_{kq}=i\left<\phi_k(t)\left|\right.\dot\phi_q(t)\right>, \qquad
 \D^{(m)}(t), \ D^{(m)}_{k'q'}=i\left<\psi_{k'}(t)\left|\right.\dot\phi_{q'}(t)\right>. 
\eeq
The matrices $\D^{(a)}(t)$ and $\D^{(m)}(t)$ are Hermitian matrices
[because the respective orbitals are normalized and 
orthogonal to one another,
$\left<\phi_k(t)\left|\right.\phi_q(t)\right>=\delta_{kq}$
and 
$\left<\psi_{k'}(t)\left|\right.\psi_{q'}(t)\right>=\delta_{k'q'}$]
and hence can be diagonalized
\beq\label{diagonal_U}
 \left\{\T^{(a)}(t)\right\}^\dag \D^{(a)}(t) \T^{(a)}(t) = \d^{(a)}(t), \qquad 
 \left\{\T^{(m)}(t)\right\}^\dag \D^{(m)}(t) \T^{(m)}(t) = \d^{(m)}(t),
\eeq
where $\d^{(a)}(t)$ and $\d^{(m)}(t)$ are the diagonal matrices of the respective eigenvalues.
Now,
we define the unitary transformations (which are {\it symbolically} integrated):
\beqn\label{equation_U}
 & & \!\!\!\!\!\!\!\!\!\! i\dot U^{(a)}_{sq}(t)= - \sum_{k=1}^M D^{(a)}_{sk}(t) U^{(a)}_{kq}(t) \ \ \Longrightarrow \ \ 
\U^{(a)}(t)=e^{+i\int^t \D^{(a)}(t') dt'} \U^{(a)}(0), \nonumber \\
 & & \!\!\!\!\!\!\!\!\!\!
i\dot U^{(m)}_{s'q'}(t)= - \sum_{k'=1}^M D^{(m)}_{s'k'}(t) U^{(m)}_{k'q'}(t) \ \ \Longrightarrow \ \
 \U^{(m)}(t)=e^{+i\int^t \D^{(m)}(t') dt'} \U^{(m)}(0), \
\eeqn
with the initial conditions defined in the limes $\tau \to 0$ as 
(see in this respect Ref.~\cite{MCTDHB2}):
\beq\label{U0_initial_conditions}
\U^{(a)}(\tau) = \T^{(a)}(0) e^{+i \tau \d^{(a)}(0)},
\qquad 
\U^{(m)}(\tau) = \T^{(m)}(0) e^{+i \tau \d^{(m)}(0)}.
\eeq
Then, 
the unitary transformations 
of the orbitals
\beqn\label{orbitals_U} 
& & \overline{\phi}_q(\r,t) = \sum_{k=1}^M U^{(a)}_{kq}(t) \phi_k(\r,t), \qquad q=1,\ldots,M, \nonumber \\
& & \overline{\psi}_{q'}(\r,t) = \sum_{k'=1}^{M'} U^{(m)}_{k'q'}(t) \psi_{k'}(\r,t), \qquad q'=1,\ldots,M' \
\eeqn
lead to the desired result -- equations of motion (\ref{MCTDH_conver_orb_eqs}) 
and (\ref{MCTDH_W_conver_orb_eqs}) --
where the projection operators 
$\hat{\mathbf P}^{(a)}$ and $\hat{\mathbf P}^{(m)}$
have been eliminated from the left-hand sides.

The transformation $\{C_{\vec{n}^p\vec{m}^p}(t)\} \to \{\overline{C}_{\vec{n}^p\vec{m}^p}(t)\}$
accompanying Eq.~(\ref{orbitals_U}), 
carries 
equations of motion (\ref{MCTDH_conver_coeff_eqs_P})
for the expansion coefficients
to the respective final result, Eqs.~(\ref{MCTDH_conver_coeff_eqs}) 
and (\ref{MCTDH_W_conver_coeff_eqs}).
It is instructive to obtain this result 
by proving that the equations of motion for the expansion
coefficients are {\it form-invariant}.
Namely,
if 
${\bcalH}^{(2a \leftrightharpoons m)}(t)\C(t) = i\frac{\partial \C(t)}{\partial t}$ 
are satisfied for the untransformed quantities 
$\left[\{C_{\vec{n}^p\vec{m}^p}(t)\}, \left\{\phi_k(\r,t)\right\}, 
\left\{\psi_{k'}(\r,t)\right\}\right]$
then 
$\overline{{\bcalH}}^{(2a \leftrightharpoons m)}(t)\overline{\C}(t) = i\frac{\partial \overline{\C}(t)}{\partial t}$ 
are satisfied for the transformed ones 
$\left[\{\overline{C}_{\vec{n}^p\vec{m}^p}(t)\}, \left\{\overline{\phi}_k(\r,t)\right\}, 
\left\{\overline{\psi}_{k'}(\r,t)\right\}\right]$.
The proof is straightforward.
Equating the variation
of the functional action 
(\ref{action_functional_AM_full},\ref{expectation_ham_b})
with respect to the expansion 
coefficients to zero,
the result can be written as follows:
$\left<\vec{n}^p,\vec{m}^p;t\left|\hat H^{(2a\leftrightharpoons m)} - 
i\frac{\partial}{\partial t}\right|\Psi(t)\right>, \forall p,\vec{n}^p,\vec{m}^p$.
Since, 
the transformed configurations $\left\{\overline{\left<\vec{n}^p,\vec{m}^p;t\right|}\right\}$ 
are given as 
linear combinations of the untransformed configurations
$\left\{\left<\vec{n}^p,\vec{m}^p;t\right|\right\}$,
the operator $\hat H^{(2a\leftrightharpoons m)} - i\frac{\partial}{\partial t}$
does not depend on the orbitals,
and $\left|\overline{\Psi}(t)\right>=\left|\Psi(t)\right>$,
we immediately get:
$\overline{\left<\vec{n}^p,\vec{m}^p;t\right|}\hat H^{(2a\leftrightharpoons m)} - 
i\frac{\partial}{\partial t}\left|\overline{\Psi}(t)\right>, \forall p,\vec{n}^p,\vec{m}^p$,
which concludes our proof.
To our needs,
since the transformed orbitals (\ref{orbitals_U}) 
obey the differential conditions
$\left<\overline{\phi}_k\left|\right.\dot{\overline{\phi}}_q\right> = 0; k,q=1,\ldots,M$
and
$\left<\overline{\psi}_{k'}\left|\right.\dot{\overline{\psi}}_{q'}\right> = 0; k',q'=1,\ldots,M'$
[see Eq.~(\ref{MCTDHI_mix_conv_const})],
the respective 
equations of motion for 
the transformed coefficients
boil down to 
$\overline{{\H}}^{(2a \leftrightharpoons m)}(t)\overline{\C}(t) = i\frac{\partial \overline{\C}(t)}{\partial t}$ 
[see Eqs.~(\ref{MCTDH_conver_coeff_eqs}) 
and (\ref{MCTDH_W_conver_coeff_eqs})].

\section{Matrix elements with multiconfigurational wavefunctions in systems with particle conversion}\label{matrix_Appen}

There are two types of matrix elements in the theory.
The first type are matrix elements of the many-body Hamiltonian with respect to
the configurations.
These matrix elements are expressed using the matrix 
elements of the one-body terms, two-body interaction terms,
and the conversion term
with respect to the 
atomic and molecular 
orbitals.
The second type of matrix elements are the 
matrix elements of the 
reduced density matrices appearing in the theory,
which are expressed 
in terms of 
the
expansion coefficients. 

In this appendix we prescribe these matrix elements.
It is easy to connect the matrix elements of
particle-conserving operators to
the corresponding matrix elements appearing
in the available multiconfigurational 
theories for identical particles and mixtures.
This assignment 
will shorten substantially the discussion below.
The matrix elements of particle non-conserving operators
are new 
and 
will be presented in full
details. 

\subsection{Matrix elements of the Hamiltonian}

The many-body 
Hamiltonian (\ref{ham_am_1}-\ref{ham_am_3})
is written as a sum of particle-conserving and particle non-conserving parts:
$\hat H^{(2a\rightleftharpoons m)} = \hat H^{(am)} + \hat W^{(2a\rightleftharpoons m)}$.
The matrix elements of the particle-conserving part $\hat H^{(am)}$, 
see Eqs.~(\ref{ham_am_1},\ref{ham_am_2}),
between two general configurations 
derive 
from the following relation:
\beq\label{ham_particle_consere_ME}
\left<\vec{n}^p,\vec{m}^p;t\left|\hat H^{(am)} - i\frac{\partial}{\partial t}\right|\vec{n}'^{p'},\vec{m}'^{p'};t\right> =
\delta_{p,p'} 
\left<\vec{n}^p,\vec{m}^p;t\left|\hat H^{(am)} - i\frac{\partial}{\partial t}\right|\vec{n}'^{p},\vec{m}'^{p};t\right>.
\eeq 
Thus, 
it corresponds to a matrix element of a mixture with
$N-2p$ atoms and $p$ molecules {\it without conversion}.
The matrix elements of the Hamiltonian of a 
mixture 
of two kinds of bosons 
with respect to two general configurations have been prescribed within 
the respective particle-conserving multiconfigurational theory for
Bose-Bose mixtures, 
the MCTDH-BB theory, 
see Ref.~\cite{MIX}.

To evaluate the matrix elements of the particle non-conserving part of the Hamiltonian,
$\hat W^{(2a\rightleftharpoons m)}=\hat W^{(2a\rightharpoonup m)} + \hat W^{(m\rightharpoondown 2a)}$,
see Eqs.~(\ref{ham_am_1},\ref{ham_am_3}),
between two general configurations we have to introduce
a shorthand notation for different configurations.
The reference configuration is denoted by
$\left|\vec{n}^p,\vec{m}^p;t\right> = 
\left|n_1^p,\ldots,n_k^p,\ldots,n_q^p,\ldots,n_M^p:m_1^p,\ldots,m_{k'}^p,\ldots,m_{M'}^p;t\right>$.
We remind that the occupation 
numbers satisfy the relations:
$|\vec{n}^p|=n_1^p+\ldots+n_M^p=N-2p$ 
and $|\vec{m}^p|=m_1^p+\ldots+m_{M'}^p=p$,
where $p$ is the number of molecules.
Now,
the configuration   
$\left|\vec{n}^{p-1}_{kq},\vec{m}^{p-1}_{k'};t\right> \equiv 
\left|n_1^p,\ldots,n_k^p+1,\ldots,n_q^p+1,\ldots,n_M^p:m_1^p,\ldots,m_{k'}^p-1,\ldots,m_{M'}^p;t\right>$
differs from $\left|\vec{n}^p,\vec{m}^p;t\right>$
by having $p-1$ molecules and $N-2p+2$ atoms,
where a molecule in the $k'$-th orbital has
dissociated to two atoms,
one in the $k$-th orbital and the second in the $q$-th orbital;
and the configuration
$\left|\vec{n}^{p-1}_{kk},\vec{m}^{p-1}_{k'};t\right> \equiv 
\left|n_1^p,\ldots,n_k^p+2,\ldots,n_M^p:m_1^p,\ldots,m_{k'}^p-1,\ldots,m_{M'}^p;t\right>$
differs from $\left|\vec{n}^p,\vec{m}^p;t\right>$
by having $p-1$ molecules  and $N-2p+2$ atoms,
where a molecule in the $k'$-th orbital has
dissociated to two atoms, 
both in the $k$-th orbital.
We employ a nomenclature where the
same ordering of the orbitals
$\phi_1,\ldots,\phi_M$ and $\psi_1,\ldots,\psi_{M'}$
as in Eq.~(\ref{MCTDH_AM_Psi}) is kept for all configurations.
In this nomenclature the following states
are equivalent:
$\left|\vec{n}^{p-1}_{kq},\vec{m}^{p-1}_{k'};t\right> \equiv 
\left|\vec{n}^{p-1}_{qk},\vec{m}^{p-1}_{k'};t\right>$.

With this notation,
the non-vanishing matrix elements of the particle non-conserving part of the Hamiltonian follow from
\beqn\label{ham_particle_NON_consere_ME}
& & \left<\vec{n}^p,\vec{m}^p;t\left|\hat W^{(2a\rightharpoonup m)}\right|\vec{n}^{p-1}_{kq},\vec{m}^{p-1}_{k'};t\right> =
 \frac{1}{\sqrt{2}} W^{(2a\rightharpoonup m)}_{k'kq} \sqrt{m_{k'}^p(n_k^p+1)(n_q^p+1)}\ , \qquad k<q, \nonumber \\
& & \left<\vec{n}^p,\vec{m}^p;t\left|\hat W^{(2a\rightharpoonup m)}\right|\vec{n}^{p-1}_{kk},\vec{m}^{p-1}_{k'};t\right> =
 \frac{1}{\sqrt{2}} W^{(2a\rightharpoonup m)}_{k'kk} \sqrt{m_{k'}^p(n_k^p+1)(n_k^p+2)}\ ,
\eeqn
and the relation 
$\left<\vec{n}^p,\vec{m}^p;t\left| \hat W^{(m\rightharpoondown 2a)}\right|\vec{n}'^{p'},\vec{m}'^{p'};t\right>=
\left\{\left<\vec{n}'^{p'},\vec{m}'^{p'};t\left| \hat W^{(2a\rightharpoonup m)}\right|\vec{n}^p,\vec{m}^p;t\right>\right\}^\ast$.
We note that 
$W^{(2a\rightharpoonup m)}_{k'kq} = W^{(2a\rightharpoonup m)}_{k'qk}$
because 
$\hat W^{(2a\rightharpoonup m)}(\r,\r') = \hat W^{(2a\rightharpoonup m)}(\r',\r)$.
To summarize,
direct coupling in the matrix
representation of the Hamiltonian
with respect to the configurations
exists due to the
particle non-conserving part of the Hamiltonian
$\hat W^{(2a\rightleftharpoons m)}$
between configurations 
with 
$p$ and $p-1$ molecules only,
for $p=1,\ldots,\left[N/2\right]$.

\subsection{Matrix elements of reduced density matrices}

The multiconfigurational ansatz (\ref{MCTDH_AM_Psi}) can be written in the following form,
\beq\label{multi_psi_appen}
\left|\Psi(t)\right> = \sum_{p=0}^{\left[N/2\right]} \left|\Psi_p(t)\right>,
\eeq
where each
$\left|\Psi_p(t)\right> = \sum_{\vec{n}^p,\vec{m}^p} 
C_{\vec{n}^p\vec{m}^p}(t) \left|\vec{n}^p,\vec{m}^p;t\right>$
is a 
(non-normalized) many-particle wavefunction 
with a {\it definite} number of 
$p$ molecules and $N-2p$ atoms,
and thus ``describes'' 
a bosonic mixture
{\it without conversion}.
Consequently,
the matrix elements of the particle-conserving 
reduced density matrices can be expressed as follows:
\beqn\label{matrix_elements_rho_conserved}
& & \rho_{kq}^{(a)}(t) = \left<\hat b_k^\dag \hat b_q\right> = 
 \sum_{p=0}^{\left[N/2\right]} \left<\Psi_p(t)\left|\hat b_k^\dag \hat b_q\right|\Psi_p(t)\right>, \nonumber \\
& & \rho_{k'q'}^{(m)}(t) = \left<\hat c_{k'}^\dag \hat c_{q'}\right> = 
 \sum_{p=0}^{\left[N/2\right]} \left<\Psi_p(t)\left|\hat c_{k'}^\dag \hat c_{q'}\right|\Psi_p(t)\right>, \nonumber \\
& & \rho_{kslq}^{(a)}(t) = \left<\hat b_k^\dag \hat b_s^\dag \hat b_l \hat b_q\right> =
 \sum_{p=0}^{\left[N/2\right]} 
\left<\Psi_p(t)\left|\hat b_k^\dag \hat b_s^\dag \hat b_l \hat b_q\right|\Psi_p(t)\right>, \nonumber \\
& & \rho_{k's'l'q'}^{(m)}(t) = \left<\hat c_{k'}^\dag \hat c_{s'}^\dag \hat c_{l'} \hat c_{q'}\right> =
 \sum_{p=0}^{\left[N/2\right]} 
\left<\Psi_p(t)\left|\hat c_{k'}^\dag \hat c_{s'}^\dag \hat c_{l'} \hat c_{q'}\right|\Psi_p(t)\right>, \nonumber \\
& & \rho_{kk'qq'}^{(am)}(t) = \left<\hat b_k^\dag \hat b_q \hat c_{k'}^\dag \hat c_{q'}\right> =
\sum_{p=0}^{\left[N/2\right]} 
\left<\Psi_p(t)\left|\hat b_k^\dag \hat b_q \hat c_{k'}^\dag \hat c_{q'}\right|\Psi_p(t)\right>. \
\eeqn
In other words,
the matrix elements of the particle-conserving reduced density matrices
can be readily read from the reduced density matrices of the 
respective particle-conserving multiconfigurational theory for
Bose-Bose mixtures, 
the MCTDH-BB theory, 
see Ref.~\cite{MIX}.

The matrix elements of the 
particle non-conserving reduced density 
matrices are given explicitly by
\beqn\label{matrix_ele_particle_non_den_gen}
 & & \rho^{(2a\rightharpoonup m)}_{k'kq} = 
\left<\hat c_{k'}^\dag \hat b_k \hat b_q\right> =
\sum_{p=0}^{[N/2]} \sum_{\vec{n}^p,\vec{m}^p} 
C^\ast_{\vec{n}^p,\vec{m}^p} C_{\vec{n}^{p-1}_{kq},\vec{m}^{p-1}_{k'}} 
\sqrt{m_{k'}^p(n_k^p+1)(n_q^p+1)}\ , \qquad k<q, \nonumber \\
 & & \rho^{(2a\rightharpoonup m)}_{k'kk} =
\sum_{p=0}^{[N/2]} \sum_{\vec{n}^p,\vec{m}^p} 
C^\ast_{\vec{n}^p,\vec{m}^p} C_{\vec{n}^{p-1}_{kk},\vec{m}^{p-1}_{k'}} 
\sqrt{m_{k'}^p(n_k^p+1)(n_k^p+2)}\ . \
\eeqn
All other 
matrix elements 
are derived
from the symmetry of the conversion 
operator,
$\rho^{(2a\rightharpoonup m)}_{k'qk}=\rho^{(2a\rightharpoonup m)}_{k'kq}$,
and the Hermiticity relation 
$\rho^{(m\rightharpoondown 2a)}_{qkk'}(t) = 
\left\{\rho^{(2a\rightharpoonup m)}_{k'kq}(t)\right\}^\ast$.


\end{document}